\documentclass[aps,jap,superscriptaddress,twocolumn,showpacs]{revtex4-1}
\usepackage{amsmath} % AMS Math Package
\usepackage{amsthm} % Theorem Formatting
\usepackage{amssymb}	% Math symbols such as \mathbb
\usepackage{graphicx} % Allows for eps images
\usepackage[usenames,dvipsnames]{pstricks}
\usepackage{pst-grad} % For gradients
\usepackage{pst-plot} % For axes
\usepackage{color}
\newcommand{\markup}[1]{{\color{Black}{#1}}}
\renewcommand{\v}[1]{\ensuremath{\mathbf{#1}}} % for vectors
\newcommand{\ket}[1]{\left| #1 \right>} % for Dirac bras
\newcommand{\bra}[1]{\left< #1 \right|} % for Dirac kets
\let\baraccent=\= % rename builtin command \= to \baraccent
\renewcommand{\=}[1]{\stackrel{#1}{=}} % for putting numbers above =

\theoremstyle{definition}

\theoremstyle{remark}


\begin{document}
%\preprint{}
\title{Quantum Confinement in Si and Ge Nanostructures}
\author{E.G. Barbagiovanni}
\email[]{ebarbagi@uwo.ca}
\affiliation{Department of Physics and Astronomy, The University of Western Ontario, London, Ontario, Canada, N6A 3K7}
\author{D.J. Lockwood}
\affiliation{National Research Council Ottawa, Ontario, Canada, K1A 0R6}
\author{P.J. Simpson}
\affiliation{Department of Physics and Astronomy, The University of Western Ontario, London, Ontario, Canada, N6A 3K7}
\author{L.V. Goncharova}
\affiliation{Department of Physics and Astronomy, The University of Western Ontario, London, Ontario, Canada, N6A 3K7}

\date{\today}

\begin{abstract}  
We apply perturbative effective mass theory as a broadly applicable theoretical model for quantum confinement (QC) in all Si and Ge nanostructures including quantum wells (QWs), wires (Q-wires) and dots (QDs). Within the limits of strong, medium, and weak QC, valence and conduction band edge energy levels (VBM and CBM) were calculated as a function of QD diameters, QW thicknesses and Q-wire diameters. Crystalline and amorphous quantum systems were considered separately. Calculated band edge levels with strong, medium and weak QC models were compared with experimental VBM and CBM reported from X-ray photoemission spectroscopy (XPS), X-ray absorption spectroscopy (XAS) or photoluminescence (PL). Experimentally, the dimensions of the nanostructures were determined directly, by transmission electron microscopy (TEM), or indirectly, by x-ray diffraction (XRD) or by XPS. We found that crystalline materials are best described by a medium confinement model, while amorphous materials exhibit strong confinement regardless of the dimensionality of the system. Our results indicate that spatial delocalization of the hole in amorphous versus crystalline nanostructures is the important parameter determining the magnitude of the band gap expansion, or the strength of the quantum confinement. In addition, the effective masses of the electron and hole are discussed as a function of crystallinity and spatial confinement.
\end{abstract}
\pacs{73.21.-b,73.22.-f,78.67.-n,61.46.-w,81.07.-b}

\maketitle

\section{Introduction\label{intro}}
Semiconductor nanostructures (NSs) exhibit increased oscillator strength due to electron hole wave function overlap, and band gap engineering due to the effect of quantum confinement (QC). Thus, materials like Si are a viable option for opto-electronics, photonics, and quantum computing.\cite{edit:2010,Lockwood:2004,Loss:1998} QC is defined as the modification in the free particle dispersion relation as a function of a system's spatial dimension.\cite{Yoffe:2002} If a free electron is confined within a potential barrier, a shift in the band gap energy is observed, which is inversely proportional to the system size squared, in the effective mass approximation. As a result, the emitted photon energy is directly proportional to the gap energy ($E_G$). QC often manifests itself in optical experiments when the dimension of the system is systematically reduced and an increase in the absorbed/emitted photon energy is measured corresponding to electron transitional states, i.e. in semiconductor NSs.

For practical applications, utilizing QC effects in NSs requires an understanding of the band structure of a low-dimensional material, how the method of preparation effects the final properties of the NS, and the kinetics/ dynamics of the absorption/emission process. The confinement potential is determined by the alignment of the respective Fermi levels when a material of a $E_{G1}$ is surrounded by a material of a $E_{G2}$, with $E_{G1}<E_{G2}$.\cite{Frensley:1977} The preparation technique can introduce stress in the system, which changes the band gap energy.\cite{Bir:1974} For indirect gap materials phonon processes can effect the recombination mechanism.\cite{Valentin:2007} The lifetime associated with the recombination event can be altered by the excitation power.\cite{VanDao:2005} (For a review of general properties of low-dimensional structures, see Refs. \onlinecite{Heiss:2005,Lockwood:2004,Yoffe:2002}. For a discussion of other higher order effects in NSs, see Ref. \onlinecite{Heiss:2005}.) This article is concerned with the electron/hole recombination process in amorphous (a) versus crystalline (c) NSs with different dimensions.

Several theoretical models (e.g. see Refs. \onlinecite{Zunger:2001,Tran:1990,Tit:2010}) have been applied to NS; all models are empirical and no one model can model all semiconductor NSs. Since the parameters of a NS system are dependent upon the preparation method for a particular material, a comprehensive theoretical understanding must test along this dimension as well. In this article, we consider a relatively simple model of direct e-h recombination using a `particle in a box' type model as a perturbation to the effective mass theory. We use no adjustable parameters \cite{Note1} and include corrections to the model dependent on the preparation method as known experimentally and/or computationally when needed, thus achieving transparency in the physics involved. The only parameter tested in this work is the crystallinity, which is shown to effect the strength of confinement (defined in Sec. \ref{theory}), because of the different symmetry properties of the electron and hole. 

The model is applied to experimental results on crystalline and amorphous Si and Ge NSs, including quantum wells (QWs), wires (Q-wires) and dots (QDs). Systems of regular shape are chosen to ensure crystallinity is the primary parameter.  For example, data obtained by van Buuren et al. \cite{Buuren:1998} for high quality `star-shaped' samples are difficult to analyse theoretically. Parameters relevant to a particular system are discussed and energy corrections are given when needed. \markup{Briefly, we compare a few theoretical models with experiment, thus, illustrating the need to categorically understand experimental parameters.} Results are discussed and a mechanism for the differences between the strength of confinement in the amorphous and crystalline system is proposed. 

\section{Theory\label{theory}}
In this work, we use the effective mass approximation (EMA) based on the Bloch periodic function. The essential features of the model are discussed below.

The Bohr radius of an electron (e), hole (h) or exciton (X) is given by, in SI units:
$$
a_{e(h)(X)}=\frac{4\pi\epsilon\hbar^2}{m^*_{e(h)(X)}e^2},
$$
$m^*_{e(h)(X)}$ is the effective mass of the e, h or X, respectively, $e$ is the electric charge and $\epsilon$ is the dielectric constant. Depending on the e or h effective mass, the X-Bohr radius is 4.5 nm for Si and 24 nm for Ge. The Bohr radius defines the spatial dimension of the particles, which determines the range of sizes for which QC can be observed. We define three regimes of confinement here:\cite{Yoffe:2002}
\begin{itemize}
\item Weak confinement: When the dimension of the system is much larger than $a_e$ and $a_h$. In this situation, the appropriate mass in the kinetic term is $M=m^*_e+m^*_h$. The energy term is dominated by the Coulomb energy. 
\item Medium confinement: When the dimension of the system is much smaller than $a_e$, but larger than $a_h$, then only electrons will experience confinement. The relevant mass is simply $m_e^*$ for the kinetic term. Most materials belong to this class.
\item Strong confinement: When the dimension of the system is much smaller than $a_e$ and $a_h$. Here both electrons and holes experience confinement and the relevant mass is the reduced mass, $\mu$, with $\frac{1}{\mu}=\frac{1}{m^*_e}+\frac{1}{m^*_h}$. In this regime, the Coulomb term is small and can generally be treated as a perturbation. 
\end{itemize}
Below we will use the terms `weak,' `medium' and `strong' to refer to the different regimes of confinement discussed above. 

Si and Ge are both indirect gap materials, meaning that, in principle, phonon scattering events are essential to maintain momentum and energy conservation during a radiative event. This situation is true in the case of a bulk material; however, as the dimension of the system is reduced, the uncertainty in the momentum $\v{k}$ vector is increased. Therefore, it is possible to break the $\v{k}$ selection rules making the $E_G$ `pseudo-direct,' allowing for direct e-h recombination.\cite{Kovalev:1998} The length scale at which this `pseudo-direct' phenomenon becomes important is typically less than a few nanometres.\cite{Hybertsen:1994,Takagahara:1992,Tit:2010} This length scale corresponds to the systems considered here; therefore, theoretically it is valid to assume direct e-h recombination without phonon-assistance. 

In the `particle in a box' model the bulk $E_G$ is taken as the ground state energy. The effect of reduced dimension is considered as a perturbation to the bulk E$_G$. Therefore, we consider the general field Hamiltonian for a system of Coulombic interacting particles given by (details are given in Ref. \onlinecite{Barbagiovanni:2011}):
\begin{eqnarray}\label{eq1}
\mathcal{H}&=&\int d^3r \psi^{\dagger}(r)\left(\frac{-\hbar^2}{2m}\bigtriangledown^2\right)\psi(r)+\notag\\
&&\frac{1}{2}\int d^3rd^3\acute{r}\psi^{\dagger}(r)\psi^{\dagger}(\acute{r})\frac{e^2}{4\pi\epsilon|r-\acute{r}|}\psi(\acute{r})\psi(r),
\end{eqnarray}
where $\psi(r)$ is the field operator, $m$ is the mass of the electron or hole, $\epsilon$ is the dielectric constant of the surrounding medium and $e$ is the electric charge. We do not consider the spin-orbit interaction here, because the fine structure is negligible at the energies considered here. 

The field operators are expanded in a two-band model for the conduction band $C$ and the valence band $V$ as:
\begin{equation}\label{eq2}
\psi(r)=\sum_k a_{k,i}\varphi_{k,i}(r)\; \; (i\in C,V),
\end{equation}
where $k$ represents a summation over momentum states. The $\varphi_{k,i}(r)$ basis set in Eq. \eqref{eq2} is expanded to reflect the use of an infinite confinement potential with a Bloch basis $u_{k,i}$. Infinite confinement is a reasonable assumption for the systems we are considering, because the matrix material has a $E_G$ several eV higher than the nano-structure; however, we can not discuss hopping or other such higher order effects. Bloch states reflect the periodic nature of the crystal (Luttinger-Kohn representation), while the boundary conditions of a NS do not reflect this same periodicity. However, in many NSs the transitions we are interested in happen near the Brillouin zone centre, e.g. the $\Gamma$-point. This statement may not be strictly true in the case of weak confinement, because $k$-selection rules are not as strongly broken as in the case of strong confinement. Nonetheless, $\v{k}\cdot\v{p}$ perturbation theory considers expansions about the Brillouin zone minimum, $\v{k}_o$. Therefore, we may justify the use of Bloch states through the use of the slowly varying wave approximation whereby only the $\v{k}_o$=0 states are retained. 

For indirect gap materials the exciton is Wannier-like, in the limit $k\ll\frac{\pi}{a_c}$ ($a_c$ is the lattice spacing) and we can drop the exchange term, which goes to zero quickly. Equation \eqref{eq1} is solved in the exciton basis using the state $\Phi$ defined as an e-h pair above the ground state, $\Phi_0$, as: $\Phi=\sum_{k_{1}k_{2}}C_{k_{1}k_{2}}a_{k_{1}}^{\dagger}b_{k_{2}}^{\dagger}\Phi_V,\;\&\;
\Phi_V =b_{k_{3}}b_{k_{4}}\cdots b_{k_{N}}\Phi_0,$ where $a_k$ ($b_k$) refers to electrons (holes) in the conduction (valence) band. Expanding in low lying $k$-states near the band edge, we solve $E_G(D)=\bra{\Phi}\mathcal{H}\ket{\Phi}$, which gives the variation of gap energy with nano-structure size.

For the mass terms in Eq. \eqref{eq1}, we use the effective masses calculated using the density of states.\cite{Yu:2001} The effective mass is related to the parabolicity of the band structure, which is not expected to change in a nano-structure compared to a bulk material at the $\Gamma$-point. Therefore, we assume the effective mass from the bulk system. For Si the effective masses at room temperature are: $m_c\rightarrow m^*_c=1.08m_o$ and $m_V\rightarrow m^*_V=0.57m_o$. For Ge the effective masses are: $m_c\rightarrow m^*_c=0.56m_o$ and $m_V\rightarrow m^*_V=0.29m_o$. These definitions yield the equation:
\begin{equation}\label{eq3}
E_{Gap}(D)=E_{Gap}(\infty)+\frac{A}{D^2}\,\text{eV}\cdot\text{nm}^2.\\
\end{equation}
$E_{Gap}(\infty)$ is the band gap of the bulk material and $D$ represents the QD diameter, the QW thickness or the Q-Wire diameter in what follows. The calculation was carried out for confinement in 1D, 2D with cylindrical coordinates and 3D with spherical coordinates. The parameter $A$ is given for Si and Ge in the strong, medium and weak confinement regimes in Table \ref{tbl1}. \markup{The change in energy of the CBM ($\Delta$E$_{CBM}$) due to QC is labelled as `medium confinement' in Table \ref{tbl1}, because a $\Delta$E$_{CBM}$ is equivalent to QC of the electron only as defined by our model, where only the electron mass is considered in Eq. \eqref{eq1}. The change in energy of the VBM ($\Delta$E$_{VBM}$) due to QC is also listed in Table \ref{tbl1}, which is calculated by considering confinement of the hole only, where only the hole mass is considered in Eq. \eqref{eq1}.} The other fixed parameter is the appropriate $E_G(\infty)$ of the bulk system and one could argue for the use of a renormalized effective mass with dimension of the system, which is discussed in Sec. \ref{disc}.

\begin{table}
\caption{Parameter $A$  given in Eq. \eqref{eq3} for 3D, 2D, 1D confinement and for $\Delta$E$_{CBM}$, $\Delta$E$_{VBM}$. \label{tbl1}}
\begin{ruledtabular}
\begin{tabular}{c c c c} 
{}& {} & {Si} & {Ge}\\
\hline
{3D}&{Strong}&{3.57}&{7.88}\\
{}&{Medium ($\Delta$E$_{CBM}$)}&{1.39}&{2.69}\\
{}&{Weak}&{0.91}&{1.77}\\
{}&{$\Delta$E$_{VBM}$}&{-2.64}&{-5.19}\\
\hline
{2D}&{Strong}&{2.09}&{4.62}\\
{}&{Medium ($\Delta$E$_{CBM}$)}&{0.81}&{1.58}\\
{}&{Weak}&{0.53}&{1.04}\\
{}&{$\Delta$E$_{VBM}$}&{-1.55}&{-3.04}\\
\hline
{1D}&{Strong}&{0.89}&{1.97}\\
{}&{Medium ($\Delta$E$_{CBM}$)}&{0.35}&{0.67}\\
{}&{Weak}&{0.23}&{0.44}\\
{}&{$\Delta$E$_{VBM}$}&{-0.66}&{-1.30}
\end{tabular}
\end{ruledtabular}
\end{table}

Finally, it is important to note that theoretical modelling can be further complicated by the accuracy of NS size determination. Transmission electron microscopy (TEM) is the direct method to determine NS size; however, if the contrast between the matrix and the nano-structure is poor, then the size uncertainty  can be on the order of 1 nm.\cite{Garrido:2004} Indirect size determinations can be used as well, such as with x-ray diffraction (XRD) \cite{Takagi:1990} or x-ray photo-emission spectroscopy (XPS).\cite{Lu:2002} Furthermore, QC in Ge has been a greater challenge for researchers to observe than in Si, because of the tendency to form defects, interfacial mixing and sub-oxide states.\cite{Barbagiovanni:2011:1,Wang:1994:1,Jin:1999,Min:1996:1} Therefore, only limited results on Ge are discussed here. However, there is recent progress in this area, showing very promising results.\cite{Rowell:2009}

\section{Experiment\label{expt}}
We cite the results of several experimental works including our own from the University of Western Ontario and from the National Research Council Ottawa, in Sec. \ref{results}. The essential features of each experiment are given here. The details of the experiments can be found in the references provided.

\section{Results\label{results}}
\subsection{Silicon\label{Si}}
\subsubsection{Quantum Well\label{QW}}
\begin{figure}
\includegraphics[scale=.7]{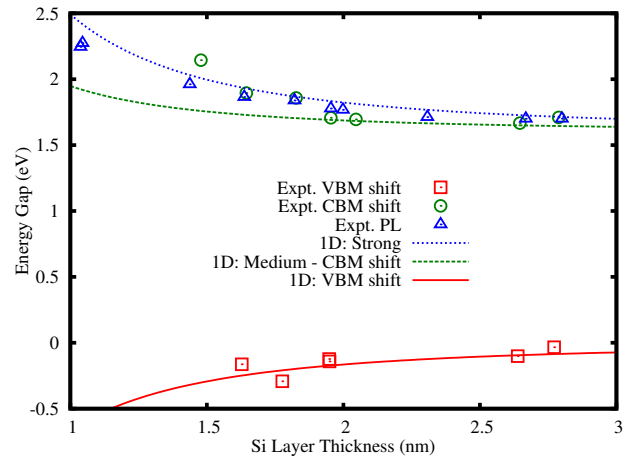} 
\caption{Disordered Si-QW data and theoretical fit. Experimental data from Ref. \onlinecite{Lu:1995}. Theoretical fit using $A$=0.89 and $E_{Gap}(\infty)=1.6$ eV in Eq. \eqref{eq3}. NB: The CBM shift is offset by the $E_{Gap}(\infty)$.\label{fig1}}
\end{figure}

Si/SiO$_2$ superlattice Si-QWs have been grown using molecular beam epitaxy, determined to be disordered via Raman scattering measurements, and their thickness found using TEM and XRD.\cite{Lu:1995, Lockwood:1996} The change in the valence band maximum (VBM) and conduction band minimum (CBM) position was measured using XPS and Si L$_{2,3}$ edge absorption spectroscopy, respectively, and room temperature PL spectroscopy was measured. Fig. \ref{fig1} plots the model predictions with the experimental data. 
 
In Ref. \onlinecite{Lu:1995} the authors used a fitting procedure according to the effective mass theory for the $\Delta$E$_{VBM(CBM)}$, resulting in $\Delta$E$_{VBM}=-0.5/D^2$ and $\Delta$E$_{CBM}=0.7/D^2$, where $D$ is the thickness of the QW. Our model predicts $\Delta$E$_{VBM}=-0.66/D^2$ and $\Delta$E$_{CBM}=0.35/D^2$. The trend for $\Delta$E$_{CBM}$ is more accurately given in Ref. \onlinecite{Lu:1995}. In Ref. \onlinecite{Lockwood:1996}, the change in $E_G$ was fitted with $A=0.7$ and $E_{Gap}(\infty)$=1.6 eV, as in Eq. \eqref{eq3}. The fit also determined the effective mass to be m$^*_{h(e)}\approx 1$. The model uses $E_{Gap}(\infty)$=1.6 eV to fit the experimental PL data well when employing the curve for strong confinement with $A=0.89$. 
 
Next we look at c-Si/SiO$_2$ QWs fabricated by chemical and thermal processing of silicon-on-insulator wafers.\cite{Lu:2002} The same methods described above were used to determine experimentally the $\Delta$E$_{VBM(CBM)}$ and the change in the gap energy including the total electron yield for a better signal to noise ratio. The thickness of the Si layer was determined by XPS using a mean free path in Si of $\sim$1.6 nm. Note that a thickness of 0.5 nm corresponds to a single unit cell of Si. Therefore, experimental data below $\approx$ 1 nm should be treated with caution. In a parallel study, these c-Si/SiO$_2$ QWs were investigated optically.\cite{Lockwood:2003}
\begin{figure}
\includegraphics[scale=.7]{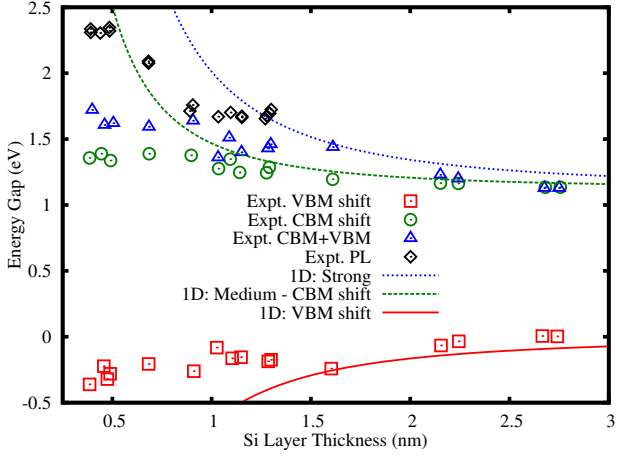} 
\caption{Crystalline Si-QW data and theoretical fit. Experimental data from Ref. \onlinecite{Lu:2002}. Experimental PL data from Ref. \onlinecite{Lockwood:2003}. Theoretical fit using $A$=0.35 and $E_{Gap}(\infty)=1.12$ eV in Eq. \eqref{eq3}. NB: The CBM shift is offset by the $E_{Gap}(\infty)$.\label{fig2}}
\end{figure}

Fig. \ref{fig2} compares experimental measurements and the model results for c-Si-QWs. The $E_G(\infty)$ in the model is 1.12 eV and the $\Delta$E$_{VBM}$ is not significant below 1.5 nm. The $\Delta$E$_{CBM}$, $\Delta$E$_{CBM+VBM}$, and the experimental PL are all well fitted by the curve for medium confinement, with $A=0.35$. In Ref. \onlinecite{Lockwood:2003} it was found that there is a second PL peak fixed with respect to the Si layer thickness at 1.8 eV. This second peak was associated with interface states. Therefore, we can assign the experimental PL data in Fig. \ref{fig2} with direct e-h recombination modelled by medium confinement.  

\subsubsection{QDs\label{QD}}
First we consider  Si QDs formed by ion implantation in SiO$_2$ films, followed by high-temperature annealing in N$_2$ and forming gas.\cite{Mokry:2009} Ref. \onlinecite{Mokry:2009} reports the QD diameter and crystalline structure observed by TEM, and room temperature PL measurements. TEM data show a Gaussian distribution in the Si-QD diameter with depth, resulting in a stretched exponential PL dynamic.\cite{Mokry:2009,Linnros:1999} 

We compare ion-implanted Si-QDs with Si QDs in a SiO$_2$ matrix prepared by microwave plasma decomposition (MPD) creating ultrafine and densely packed Si QDs\cite{Takagi:1990} (implying that tunnelling effects are important here \cite{Kamenev:2004}). The crystallinity and size was determined by TEM imaging and XRD, respectively. In Ref. \onlinecite{Takagi:1990}, the authors note that PL was not observed unless the Si QDs were oxidized, implying that surface bonds were passivated with suboxide states eventually forming a surround SiO$_2$ matrix.

\begin{figure}
\includegraphics[scale=.7]{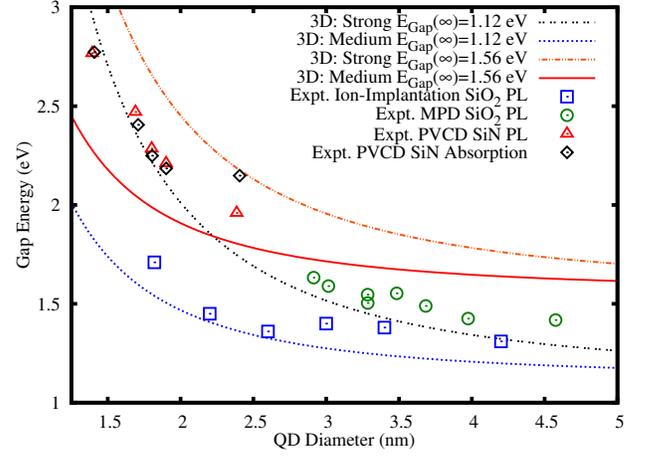} 
\caption{Crystalline and amorphous Si-QD data and theoretical fit. `Expt. Ion-Implantion SiO$_2$' refers to crystalline Si QDs embedded in SiO$_2$ from Ref. \onlinecite{Mokry:2009}. `Expt. microwave plasma decomposition (MPD) SiO$_2$' refers to crystalline Si QDs embedded in SiO$_2$ from Ref. \onlinecite{Takagi:1990}. `Expt. plasma enhanced chemical vapour deposition (PCVD) SiN' refers to amorphous Si QDs embedded in SiN from Ref. \onlinecite{Park:2001}. Theoretical fit using $A$=3.57 and 1.39 and $E_{Gap}(\infty)=1.12\; \text{or}\; 1.56$ eV (as labeled) in Eq. \eqref{eq3}. NB: The absorption data is offset by the $E_{Gap}(\infty)$.\label{fig3}}
\end{figure}

Fig. \ref{fig3} shows the experimental PL data for ion-implantated and MPD Si QDs together with our calculated curves for strong and medium confinement. Above 3 nm both sets of experimental data follow closely the model of strong confinement with $A$=3.57 and $E_G(\infty)$=1.12 eV. This indicates that for sample diameters larger than this size tunnelling effects are significant, implying a de-localization of carrier states. Iacona et al. measured a similar trend for experimental PL data. \cite{Iacona:2000} Below 3 nm, when QC effects are particularly strong, the ion-implantation data follows the curve for medium confinement, with $A$=1.39. 

Next we consider a-Si QDs embedded in a SiN matrix.\cite{Park:2001}  The Si QDs were fabricated using plasma enhanced chemical vapour deposition. The size and amorphous structure were measured using TEM and the PL was taken at room temperature. Absorption data was taken by ultraviolet-visible absorption spectroscopy. The value for the bulk band gap given by the authors is 1.56 eV, which is obtained via a fitting procedure. This value is known to vary between 1.5$\rightarrow$1.6 eV, for Si samples prepared similarly.\cite{Park:2001} 

We can see in Fig. \ref{fig3} that the experimental data for absorption and PL of a-Si QDs embedded in SiN lies between the curve for medium ($A$=1.39) and strong ($A$=3.57) confinement, with $E_{Gap}(\infty)$=1.56. Using a fitting procedure, the authors of Ref. \onlinecite{Park:2001} found $A$=2.40. The authors further conclude that by observing the fact that the experimental absorption data lies close to the PL data, one can conclude that the PL data for these samples is a good measure of the actual change in the $E_G(D)$.\cite{Park:2001} Notice that this situation is similar to that observed for Si-QWs (see Fig. \ref{fig1} and \ref{fig2}).

\subsubsection{Quantum Wires\label{QWire}}
Due to inherent complications in the fabrication process {of Si or Ge wires with a diameter below the Bohr radius,} few studies on QC in nano-wires exist and we are only able to report on c-Si-Q-wires. On the other hand, por-Si studies are widely cited in the literature. With suitable control of the etchant, por-Si QDs can become elongated,\cite{Seo:2009} thus breaking confinement in one direction implying they are more wire-like; a detailed discussion is provided in Ref. \onlinecite{Hamilton:1995}. In this case, they are called pseudo-por-Si-QDs or in the case they behave like interconnected dots, spherites.\cite{Lockwood:1995} 

Anodically grown por-Si samples were prepared by Schuppler et al.\cite{Schuppler:1994} X-ray absorption measurements determined the structures to be closer to c-Si than to a-Si. TEM was used to determine the size and PL measurements were performed at room temperature. The por-Si structures are said to be H-passivated and O-free; however, samples were exposed to air.  

Si Q-wires were produced by Ma et al. using an oxide-assisted growth method with SiO powders.\cite{Ma:2003} Subsequently, the wires were cleaned with HF to remove the oxide, thus forming a H-terminated surface. Scanning tunnelling microscopy was used to determine the diameter of the wires. The formation of SiH$_2$ and SiH$_3$ was observed on the facets of the Q-wires, which was attributed to bending stresses in the wires. The energy gap was determined using scanning tunnelling spectroscopy, which also indicated doping levels in the wires as seen by an asymmetrical shift of the $E_G$ around 0 V.  

\begin{figure}
\includegraphics[scale=.7]{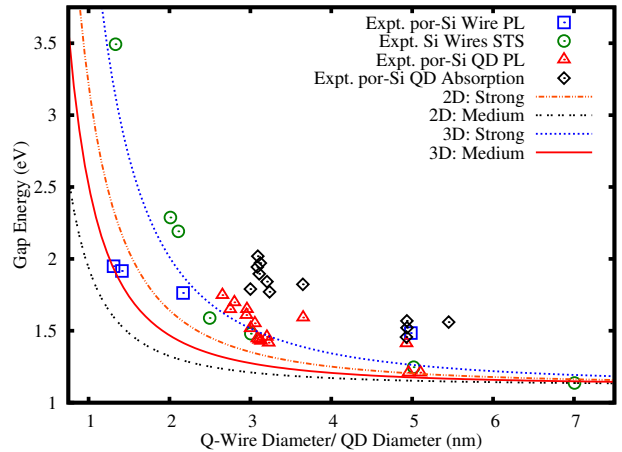} 
\caption{Crystalline Si Q-wire and QD data and theoretical fit. Experimental por-Si wire data from Ref. \onlinecite{Schuppler:1994}. Experimental Si Wire data from Ref. \onlinecite{Ma:2003}, using scanning tunnelling spectroscopy (STS). Experimental por-Si QD data from Ref. \onlinecite{Lockwood:1995, Lockwood:1994}. Theoretical fit using $A$=1.39 and 0.81 and $E_{Gap}(\infty)=1.12$ eV in Eq. \eqref{eq3}. NB: The absorption data is offset by the $E_{Gap}(\infty)$. NB: `QD' here refers to spheroids. \label{fig4}}
\end{figure}

The experimental data from Ma et al. and Schuppler et al. can be seen in Fig. \ref{fig4}. Below 3 nm the experimental data from Schuppler et al. (`por-Si Wire PL') lie close to the curve for 2D strong confinement with $A$=2.09 and $E_{Gap}(\infty)$=1.12 eV. Notice that the experimental data also lie close to the curve for 3D medium confinement with $A$=1.39. This observation may be a reflection of the idea that these structures are between dots and wires. On the other hand, the data from Ma et al. lie close to the curve for 3D strong confinement, using the same $E_{Gap}(\infty)$ and $A$=3.57. We also note that recently Si-Q-wires have been produced\cite{Yoshioka:2011} with results nearly identical to those of Ma et al. 
 
Experimental data on pseudo-por-Si-QDs for both absorption and PL are taken from Ref. \onlinecite{Lockwood:1995, Lockwood:1994}. Raman and TEM measurements were used to determine the size and the `spherite' nature of the samples, respectively. PL measurements were performed at room temperature and at 4.2K, with very little difference in the two measurements. Optical absorption was performed at room temperature. It is also noted in Ref. \onlinecite{Lockwood:1994} that, for por-Si, interface states and phonon events are significant. Fig. \ref{fig4} shows the PL and absorption experimental data for por-Si-QDs. Here the experimental data are modelled by the curve for 3D strong confinement, with $A$=3.57 and the same gap energy as above. Compared to absorption and PL data for a-Si-QDs in Fig. \ref{fig3} and the Si-QWs in Fig. \ref{fig1} and \ref{eq2}, there is a significant shift between the absorption data and the PL data, indicating a Stokes shift in the emission.\cite{Lockwood:1994} Furthermore, as noted in Ref. \onlinecite{Lockwood:1995}, the experimental PL data are nearly identical to Takagi et al., shown in Fig. \ref{fig3}.
 
\subsection{Germanium\label{Ge}}
The first observation of QC in Ge was by Takeok et al.\cite{Takeoka:1998} In this study, they produced Ge QDs using an rf co-sputtering method followed by thermal annealing. The size of the Ge QDs was controlled by varying the initial Ge concentration and was later determined by TEM imaging, which also showed that the Ge QDs were highly crystalline. PL was performed at room temperature. 

\begin{figure}
\includegraphics[scale=.7]{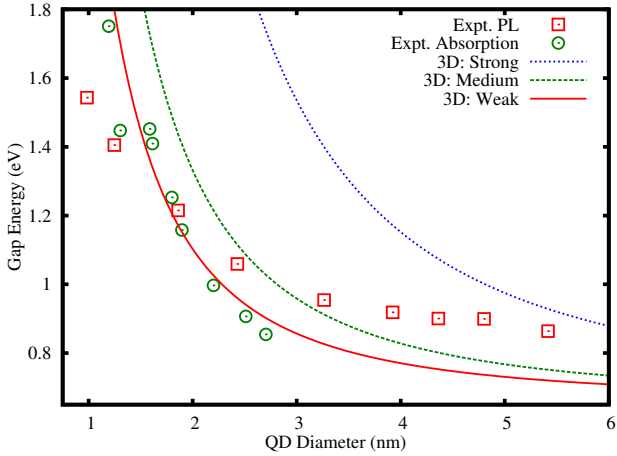} 
\caption{Crystalline Ge-QD data and theoretical fit. Experimental Ge QDs data from Ref. \onlinecite{Takeoka:1998}. Experimental Ge absorption data from Ref. \onlinecite{Bostedt:2004}. Theoretical fit using $A$=2.69 and $E_{Gap}(\infty)=0.66$ eV in Eq. \eqref{eq3}.\label{fig5}}
\end{figure}

In a more recent study, Ge QDs were produced by condensation out of the gas phase onto a Si substrate cleaned by HF.\cite{Bostedt:2004} The Ge QDs were determined to be in the bulk diamond crystalline phase. X-ray absorption (XAS) data were taken and can be seen in Fig. \ref{fig5}. XAS excites the Ge 2p electron into the conduction band; therefore, the researchers obtained data for the change in the conduction band. 

The experimental data from Refs. \onlinecite{Takeoka:1998,Bostedt:2004} are presented in Fig. \ref{fig5}. Note that the absorption data are obtained by shifting the Bostedt et. al. data by the $E_G(\infty)$ of Ge at 0.66 eV. Further note that above 3 nm there is a nearly identical departure from the medium confinement curve into strong confinement as was seen with Si-QDs in Fig. \ref{fig3}. In general, both sets of experimental data are well modelled by the curve for medium confinement with $A$=2.69 and $E_{Gap}(\infty)=0.66$ eV. For the smaller sizes (below 2.5 nm) the behaviour appears to deviate from medium confinement. This result may be because for the smaller sizes the authors only estimated the sizes.\cite{Takeoka:1998} In Ref. \onlinecite{Bostedt:2004} the Ge-QD diameter was determined using atomic force microscopy, which can potentially give a larger uncertainty in determining the size of the dot.\cite{Eaton:2010} Therefore, if the QDs are not symmetric then the diameter measurements could be inaccurate.

\section{Discussion\label{disc}}
We start our analysis by giving a justification of the EMA, while highlighting some of the limitations. Extensive arguments appear in the literature concerning the validity of the EMA and its $\v{k}\cdot\v{p}$ generalization. On the one hand, it is argued and demonstrated that the EMA overestimates the $E_G$\cite{Zunger:2001,Yoffe:2002}; however, Sec. \ref{results} demonstrated that in some cases the EMA can underestimate the $E_G$. In part, this is because due to QC the parabolic nature of the bands is possibly removed. Another complication can arise from the fact that the envelope functions may not be slowly-varying over the unit cell, which is essentially complicated by the boundary conditions. A central problem for EMA is in its applicability to a-materials, because it is based on the assumption of translational symmetry. Street has argued that while it is strictly not justified in the a-system, due to nonspecifically-defined $\v{k}$ vectors, it is still widely used albeit with differing assumptions.\cite{Street:1991} We will discuss further the application of the EMA to both amorphous and nanostructured-systems below.

On the other hand, it has been argued by S\'{e}e et al. that the EMA is well justified and produces agreement with the tight-binding method.\cite{See:2002} Such arguments reside in the fact that it is not clear what all the relevant parameters are in a nano-structured system of a particular material. In general, the boundary conditions of the system become very important, which is a problem for all theories.\cite{Lassen:2009} If the Fourier components of the envelope function are centred around the the Brillouin zone centre, then envelope functions can be justified. In addition, this justification has been extended to consider that if the interface is defect free then the EMA is justified.\cite{Lassen:2009} Other considered corrections to the EMA use a fourth order term in $\v{k}$.\cite{Yoffe:2002} The advantage of the EMA is that it is straightforward in its application, thus allowing one to highlight key features of individual systems. Perturbations in the NS system are naturally treated in the $\v{k}\cdot\v{p}$ method and defect states easily calculated\cite{Yu:2001}. Compared to empirical methods,\cite{Delerue:1993,Zunger:2001} which produce a dimensional dependence of $D^{-1.39}$, the EMA has the units $D^{-2}$ (see \eqref{eq3}). In addition, it has been shown that the $\v{k}\cdot\v{p}$ Hamiltonian can be made to reproduce multiband coupling effects and the correct symmetry of the QD.\cite{Tomic:2011}

\begin{figure}
\includegraphics[scale=.5]{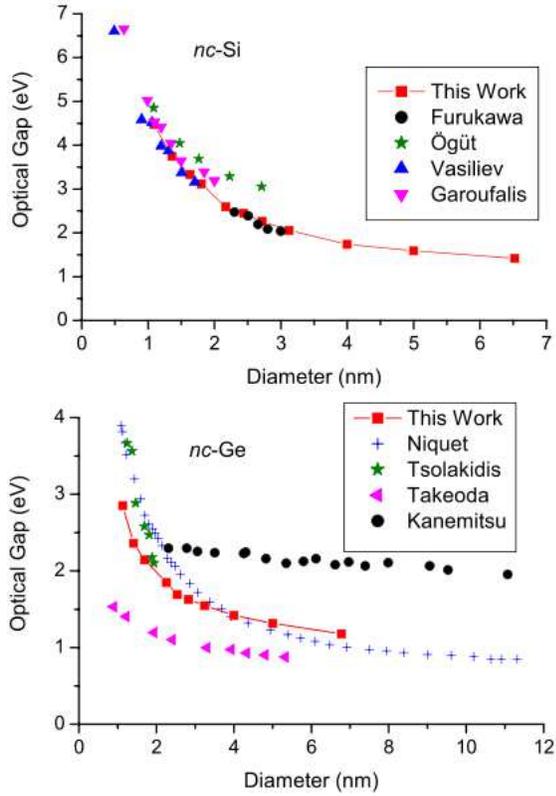} 
\caption{Comparison of the E$_G$ as a function of QD diameter for Si and Ge. Reprinted figure with permission from Ref. \onlinecite{Bulutay:2007}. Copyright (2007) by the American Physical Society. References for the works listed in figure caption are given in Ref. \onlinecite{Bulutay:2007}. \label{fig6}}
\end{figure} 

\markup{To emphasize the importance of accurately parametrizing the preparation method, we compare our results with a few theoretical models with respect to experiment. In the work by Bulutay\cite{Bulutay:2007}, the variation in the E$_G$(D) is calculated using an atomistic pseudopotential method. The result is given in Fig. 4 of Ref. \onlinecite{Bulutay:2007} and is reproduced here in Fig. \ref{fig6}. References for the works listed in figure caption are given in Ref. \onlinecite{Bulutay:2007}. The top curve in Fig. \ref{fig6} is for Si, which we compare with our Fig. \ref{fig3} here and the bottom curve for Ge, which we compare with our Fig. \ref{fig5} here. It is clear that the band gaps 
shown in Fig. \ref{fig6} are consistently larger than what we present in this 
manuscript. This result is easy to explain.

In the case of Si, the experimental data in Fig. \ref{fig6} is from Furukawa et al. In this work, they produce Si QDs using magnetron rf sputtering. It is demonstrated that the QDs are surrounding by H and composed of Si:H. The incorporation of H in the QDs causes an increase in the E$_G$ (see Fig. 1 of Ref. \cite{Furukawa:1988}). Furthermore, the Raman peak of the QDs is measured to be 514 cm$^{-1}$, as opposed to the bulk value at 520 cm$^{-1}$, which indicates that the system is under stress. To verify this claim, the authors measure a 2\% extension in the bond length using x-ray diffraction. Thus, the QDs are under tensile stress, which increases the E$_G$\cite{Hong:2003}, see below for more detail. Finally, Furukawa et al. explain that the origin of the stress is through the incorporation of the above mentioned H in the Si QD lattice through a plasma-assisted crystallization process. Whereas, if H was acting only to passivate the dangling bonds there would be no change in the E$_G$.\cite{Lopez:2002} Therefore, the experimental results of Furukawa et al. are higher than what is presented here (Fig. \ref{fig3}) because of an additional stress component, which increase that E$_G$ beyond that of QC alone and which is introduced because of the preparation method. 

The remaining results in Fig. \ref{fig6} for Si are theoretical results. The work by \"{O}\u{g}\"{u}t et al. uses a real-space pseudopotential method. Vasiliev et al. uses linear-response within the time-dependent local density approximation. Garoufalis uses time-dependent density functional theory. Including the work of Bulutay, all four methods give approximately the same result. However, looking at Bulutay, there is no explicit inclusion of a stress component in the Hamiltonian, instead it is implicitly fitted in the pseudo-potential, while the other three methods ignore stress altogether. Therefore, these four methods do not explicitly consider the experimental details, instead they are fitted to experiment. The empirical nature of the theoretical methods can be further seen when comparing with similar pseudopotential calculations that produce different results from those shown here.\cite{Zunger:2001} 

Considering the results for Ge in Fig. \ref{fig6}, the situation is essentially the same as for Si above. The experimental data from Kanemitsu et al. is associated with defect PL only, making it beyond the scope of this paper. Niquet et al. uses an sp$^3$ tight binding method, while Tsolakidis et al. uses time-dependent density functional theory in the adiabatic local density approximation. The experimental data from Takeoka et al. is fitted in this manuscript (Fig. \ref{fig5}) and not in the references of Fig. \ref{fig6}.  Furthermore, these calculations for Ge are similar to Si, which implies the material properties are not being properly accounted for theoretically. Thus, it is necessary to quantify each term in the Hamiltonian according to the preparation method.}

Finally, we comment on the relevant energy scales for the experiments considered above. The electron and hole can form a hydrogenic or positronium-like exciton, a bound state of the constituent particles, thus modifying the photon energy during the recombination event by the Coulomb interaction between the electron and hole. The Coulomb energy is on the order of hundreds of meV ($\rightarrow 1/R$, $R$=NS dimension), the exchange energy is on the order of 0.1 meV ($\rightarrow 1/R^3$), while the gap energy is on the order of several eV. Due to the large number of competing parameters in any real system, the exact value of the above parameters is not known, and these are important for precision control of a device.

To summarize the comparisons made in Sec. \ref{results}, we first consider the relationship between experimental absorption and PL data. In the case of disordered-Si-QWs (Fig. \ref{fig1}), c-Si-QWs (Fig. \ref{fig2}) and a-Si-QDs in SiN (Fig. \ref{fig3}) the absorption curve follows closely with the PL. As mentioned in Sec. \ref{QD}, this result indicates that the PL  measurement is an accurate measure of $E_G(D)$. Furthermore, in the case of Si-QWs the VBM does not change significantly. Therefore, we conclude that the model dependence between these three systems does not lie in the change in the VBM. 

Considering the absorption data from por-Si-QDs (Fig. \ref{fig4}), there is a significant shift between the absorption data and PL data, which was noted in Sec. \ref{QWire}. In addition, the por-Si QD data are nearly identical to the MPD Si-QDs (Fig. \ref{fig3}), which indicates that these systems are structurally similar with similar decay dynamics. In the case of por-Si it has been found that this system is under tensile stress.\cite{Kun:2005} Tensile stress, which is a function of the thickness of oxide, is known to increase the band gap.\cite{Hong:2003} It is known that the surrounding oxide has a strong effect on the resulting PL in por-Si.\cite{Wolkin:1999} The resulting Si-O-Si bonds due to the oxidation process place large stresses on the por-Si crystallites. In addition, it has been shown that the dominant PL comes from surface states.\cite{Lockwood:1994} At the surface or interface states, it has been shown that band bending on the order of 0.2$\rightarrow$0.3 eV can occur.\cite{Svechnikov:1998} Such a shift in energy corresponds with the discrepancy shown in Fig. \ref{fig3} and \ref{fig4}.

For the c-Si-Q-wires measured by STS (Fig. \ref{fig4}), the data are modelled by strong confinement. This is because of the stresses observed in the system and possibly because of the doping; both are factors that do change the nature of electronic structure. In Sec. \ref{QWire}, we mentioned that these structures experience bending stress, which has a tensile component. Furthermore, Fig. \ref{fig4} illustrates that c-Si-Q-wires are identical in energy to por-Si; therefore, the analysis of these systems is similar. By contrast, the por-Si wire PL data (Fig. \ref{fig4}) behaves more wire-like, which may be a result of the fact the authors took care to minimize oxygen exposure (see Sec. \ref{QWire}).

From Fig. \ref{fig3} and \ref{fig5}, both ion implanted Si-QDs and Ge-QDs have the same behaviour above 3 nm. They lie close to the curve for strong confinement, similar to the case of por-Si, indicating that possible stresses or interface states are important in this regime. Ge is known to experience stress in a SiO$_2$ matrix.\cite{Sharp:2005} Tensile stress can be relieved depending on the nature of the interface bonds and the surface to volume ratio of Si:SiO$_2$.\cite{Hong:2003} In the work of Ref. \onlinecite{Barbagiovanni:2011} it was found from Raman spectroscopy that ion-implanted QDs are not under stress for diameters smaller than 3 nm. Therefore, c-Si-QDs produced by ion implantation and c-Ge-QDs are well modelled by medium confinement below 3 nm.   

Finally, a-Si-QDs in SiN (Fig. \ref{fig3}) lie between medium and strong confinement (see Sec. \ref{QD}). SiN has a band gap of 5.3 eV versus SiO$_2$ at 9.2 eV, which allows for tunnelling of carrier states.\cite{Park:2001} More importantly, if we consider the nucleation process during thermal annealing and consider the bond enthalpies for diatomic species (SiN at 470 kJ/mol and SiO at 799 kJ/mol), it is easier to break SiN bonds, thus allowing for a greater degree of intermixing at the QD-matrix interface. Therefore, a SiN matrix acts more like a finite potential barrier, which lowers the gap energy from the infinite case. A numerical computation indicates that the difference between the case of finite versus infinite confinement potential is between 10$\%$ and 15$\%$ depending on the size of the system. This difference exactly corresponds with the difference we see in Fig. \ref{fig3}. Therefore, we conclude that a-Si-QDs in SiN are well modelled by strong confinement. 

From the results above and considering modifications that must be made to our model to account for non-direct e-h recombination phenomena, it is clear that strong confinement describes a-materials and medium confinement describes c-materials. Therefore, since QC of a particle is a function of the delocalization of that particle with respect to the dimension of the system, we need to account for the fact that the hole becomes more delocalized in the a-system than in the c-system. This fact may or may not be seen as a shift in the VBM. As noted above, disordered-Si-QWs, c-Si-QWs and a-Si-QDs in SiN all do not show a large variation in the VBM. 

A mechanism for pinning of the hole states in c-Si-QDs was discussed in the work of Sa'ar et al. as a function of the hole coupling with vibrons.\cite{Saar:2005} However, this phenomenon does not account for the fact that the hole becomes more delocalized in the a-system, it is well known that band-tail states play a very important role in the band structure of a-materials, even though the population density is relatively low.\cite{Street:1991} Kanemitsu et al. (and Refs. within), report the experimental observation that the band-tail states become strongly delocalized in the a-system, while the hole remains relatively localized in the c-system.\cite{Kanemitsu:2002} This observation accounts for what is observed in this work. 

Another critical factor to discuss is the effective mass concept, particularly in the a-system. Recall from Sec. \ref{theory}, the bulk effective mass is used in the calculations. It is possible that this parameter is not well-justified in the a-system\cite{Street:1991} and is simply not valid in the nano-structured system, in the worst possible case, or it is size-dependent.\cite{Seino:2011,Rossner:2003,Yang:2004} 

The electron (e) and hole (h) interact differently with the atomic structure. The s-like electron has $\Gamma^6_c$ symmetry; the p-like hole is contained in $\Gamma^8_v$. Therefore, holes interact more strongly with the acoustical lattice vibrations. The electron has approximately twice the mass of the hole, which is dependent on the gap energy. Hence the crystallinity will effect the properties of the particles differently and recombination events are dependent on such properties.

The a-system has typically $80\%$\cite{Ley:1984} of the density of the c-system, while disordered Si generally refers to a density $\approx98\%$\cite{Lu:1996} of that of c-Si, and these values can vary widely based on the preparation method.\cite{Renmer:1973} Therefore, short and medium range structural order does remain in both of these systems. Although the long-range order is not well-defined in the a-system along with the $\v{k}$-vectors, alternative approaches to this concept have been extensively presented. In an earlier work, Kivelson et al. defined an alternative approach to this concept.\cite{Kivelson:1979} They formulated the assumptions (i) the structure of the solid can be approximated by a rigid continuous random network that is homogeneous on the scale of the slowly varying envelope, and (ii) the band can be measured by a set of linearly independent orbits, which are not necessarily orthogonal. Furthermore, Kivelson et al. used a tight-binding approach with approximate eigenvalues to obtain the effective mass Hamiltonian. In another approach\cite{Singh:2002}, Singh looked at the effective masses in the extended and tail-states around the mobility edge directly using a real-space formulation. The electron energy eigenvalues are given in terms of the probability amplitude, which cannot be defined as in the case of a c-material in terms of $\v{k}$-vectors. Instead, the probability amplitude is defined as\cite{Singh:2002}:
$$
C_{1\v{l}}=N^{-1/2}\exp(i\v{s}_e\cdot\v{l}),\;\text{with}\;\v{s}_e(E)=\sqrt{\frac{2m_e^*(E-E_C)}{\hbar^2}}
$$
where $E_C$ defines the mobility edge; therefore, the effective mass is defined above the mobility edge in the extended states and is imaginary in the tail states. In either approach described here, the result is that the effective mass calculated is lower than in the bulk system. This observation implies that the Bohr radius of the hole in the a-material is larger than in the c-material and hence the hole is more delocalized in the a-material, thus the observed strong confinement. It is clear that this is the dominant mechanism for strong confinement in the amorphous system, since the pinning discussion above does not describe all the systems considered here. The relative magnitude of the two mechanisms needs further analysis.

The size dependence of the effective mass in c-systems is reported in Refs. \onlinecite{Seino:2011,Rossner:2003,Yang:2004}. Experimentally, the effective mass is reported to decrease with size in Ref. \onlinecite{Rossner:2003,Yang:2004}. In one theoretical report, the hole effective mass increases, while the electron effective mass decreases.\cite{Seino:2011} The magnitude of change in the effective mass is roughly the same for the electron and the hole, and considering the effective mass of the electron in the bulk system is roughly twice that of the hole, it is not likely that the change will be within experimental resolution. Overall, the effective mass in the a-system and in the nano-structured system is understood to decrease, but the magnitude of the decrease is unclear. Therefore, in terms of the calculations presented here, if the effective mass is lowered than we should expect to see an increase in the calculated $E_G$ and hence our curves will shift upwards. However, we would also expect to see an increase in $E_G$ from the experimental results. Since the exact magnitude of these changes is not known it is difficult to evaluate the error incurred by using the bulk effective mass.

\markup{This issue of the correct effective mass is more poignant when considering the $\Delta$E$_{CBM(VBM)}$. In this work, $\Delta$E$_{VBM}>\Delta$E$_{CBM}$, which is understood, because the effective mass of the hole is smaller than the electron. However, experiment consistently shows the opposite effect, see Fig. \ref{fig1},\ref{fig2} and see Ref. \onlinecite{Afanasev:2009,Seguini:2011}. This observation implies that experiment is measuring a larger decrease in the electron effective mass than the hole, or possibly a relative increase in the hole mass compared to the electron. This observation is nearly consistent with Ref. \onlinecite{Seino:2011}, where they predict a nearly symmetric change. Furthermore, recall that experiment reports a decrease in the electron effective mass.\cite{Rossner:2003,Yang:2004} 

Therefore, the decrease in the electron effective mass and increase in the hole effective mass is consistent for the crystalline system with our observation of medium confinement, because the hole is more spatially localized. In being consistent with experiment, we drop the hole contribution for the crystalline system in the ideal approximation, because this term is not as significant as the electron according to the $\Delta$E$_{CBM(VBM)}$ measurements, described above. Although, there may still be a slight hole contribution in this ideal approximation, which needs further study. In addition, in our theoretical modelling, we have consistent results for strong confinement in the amorphous samples, because both the electron and hole effective mass decrease implying confinement of both, due to spatial de-localization. Although, the relative contribution from the electron versus the hole is not clear and needs further study. Furthermore, it is clear that the effective mass prediction for $\Delta$E$_{CBM(VBM)}$ is not correct, unless a renormalized effective mass is used according to system dimension. These results are a clear indication that the use of the bulk effective mass is only a first order approximation. Nevertheless, very good overall agreement is obtained between experiment and a theory with essentially no adjustable parameters for both Si and Ge nanostructures.}

\section{Conclusion\label{conl}}
We have studied the effect of confinement dimensions and crystallinity on the magnitude of the band gap expansion (as a function of decreasing size) in group IV semiconductor NSs (quantum wells (QWs), wires (Q-wires) and dots (QDs). Medium and strong confinement models provide the best fit to experimental results; moreover crystalline materials exhibit medium confinement, while amorphous materials exhibit strong confinement regardless of the confinement dimensions of the system. This difference in confinement strength was explained by considering the extent of spatial delocalization of the hole. A possible explanation is hole pinning due to coupling with the vibronic states.\cite{Saar:2005} It has previously been reported\cite{Kanemitsu:2002} that band tail states become strongly delocalized in the amorphous system compared to the crystalline system. This hole delocalization would partially account for the trends observed in our work. The concept of the effective mass was reviewed for the amorphous system. We argue that the effective mass can still be defined in the amorphous material around the mobility edge.\cite{Kivelson:1979,Singh:2002} A lower value of the effective mass is reported for the amorphous system, which accounts for the trends observed in our work, while the hole mass increases and the electron mass decreases as a function of spatial confinement.\cite{Seino:2011,Rossner:2003,Yang:2004} With the diminished effective mass (the absolute value of this change is not possible to estimate, and more work is needed in this area), we expect an increase in $E_{Gap}$, and our calculated curves of energy versus diameter will be shifted upwards. However the general trends observed in this work will remain the same.

\bibliography{refr}

%Merlin.mbs v4.21 2009-07-09.
\begin{thebibliography}{10}%
\makeatletter
\providecommand \@ifxundefined [1]{%
 \ifx #1\undefined \expandafter \@firstoftwo
 \else \expandafter \@secondoftwo
\fi
}%
\providecommand \@ifnum [1]{%
 \ifnum #1\expandafter \@firstoftwo
 \else \expandafter \@secondoftwo
\fi
}%
\providecommand \enquote [1]{``#1''}%
\providecommand \bibnamefont  [1]{#1}%
\providecommand \bibfnamefont [1]{#1}%
\providecommand \citenamefont [1]{#1}%
\providecommand\href[0]{\@sanitize\@href}%
\providecommand\@href[1]{\endgroup\@@startlink{#1}\endgroup\@@href}%
\providecommand\@@href[1]{#1\@@endlink}%
\providecommand \@sanitize [0]{\begingroup\catcode`\&12\catcode`\#12\relax}%
\@ifxundefined \pdfoutput {\@firstoftwo}{%
 \@ifnum{\z@=\pdfoutput}{\@firstoftwo}{\@secondoftwo}%
}{%
 \providecommand\@@startlink[1]{\leavevmode\special{html:<a href="#1">}}%
 \providecommand\@@endlink[0]{\special{html:</a>}}%
}{%
 \providecommand\@@startlink[1]{%
  \leavevmode
  \pdfstartlink
   attr{/Border[0 0 1 ]/H/I/C[0 1 1]}%
   user{/Subtype/Link/A<</Type/Action/S/URI/URI(#1)>>}%
  \relax
 }%
 \providecommand\@@endlink[0]{\pdfendlink}%
}%
\providecommand \url  [0]{\begingroup\@sanitize \@url }%
\providecommand \@url [1]{\endgroup\@href {#1}{\urlprefix}}%
\providecommand \urlprefix [0]{URL }%
\providecommand \Eprint[0]{\href }%
\@ifxundefined \urlstyle {%
  \providecommand \doi [1]{doi:\discretionary{}{}{}#1}%
}{%
  \providecommand \doi [0]{doi:\discretionary{}{}{}\begingroup
  \urlstyle{rm}\Url }%
}%
\providecommand \doibase [0]{http://dx.doi.org/}%
\providecommand \Doi[1]{\href{\doibase#1}}%
\providecommand \bibAnnote [3]{%
  \BibitemShut{#1}%
  \begin{quotation}\noindent
    \textsc{Key:}\ #2\\\textsc{Annotation:}\ #3%
  \end{quotation}%
}%
\providecommand \bibAnnoteFile [2]{%
  \IfFileExists{#2}{\bibAnnote {#1} {#2} {\input{#2}}}{}%
}%
\providecommand \typeout [0]{\immediate \write \m@ne }%
\providecommand \selectlanguage [0]{\@gobble}%
\providecommand \bibinfo [0]{\@secondoftwo}%
\providecommand \bibfield [0]{\@secondoftwo}%
\providecommand \translation [1]{[#1]}%
\providecommand \BibitemOpen[0]{}%
\providecommand \bibitemStop [0]{}%
\providecommand \bibitemNoStop [0]{.\EOS\space}%
\providecommand \EOS [0]{\spacefactor3000\relax}%
\providecommand \BibitemShut [1]{\csname bibitem#1\endcsname}%
%</preamble>
\bibitem{edit:2010}%
  \BibitemOpen
  \bibfield{author}{%
  \bibinfo {author} {\bibnamefont{editorial}},\ }%
  \bibfield{journal}{%
  \bibinfo {journal} {Nature Nanotechnology}\ }%
  \textbf{\bibinfo {volume} {5}},\ \bibinfo {pages} {381} (\bibinfo {year}
  {2010})%
  \bibAnnoteFile{NoStop}{edit:2010}%
\bibitem{Lockwood:2004}%
  \BibitemOpen
  \bibfield{author}{%
  \bibinfo {author} {\bibfnamefont{D.~J.}\ \bibnamefont{Lockwood}}\ and\
  \bibinfo {author} {\bibfnamefont{L.}~\bibnamefont{Pavesi}},\ }%
  \enquote{\bibinfo {title} {Silicon photonics},}\ in\ \emph{\bibinfo
  {booktitle} {Topics in Applied Physics}},\ Vol.~\bibinfo {volume} {94},\
  \bibinfo {editor} {edited by\ \bibinfo {editor} {\bibfnamefont{D.~J.}\
  \bibnamefont{Lockwood}}\ and\ \bibinfo {editor}
  {\bibfnamefont{L.}~\bibnamefont{Pavesi}}}\ (\bibinfo {publisher} {Springer},\
  \bibinfo {address} {Berlin},\ \bibinfo {year} {2004})\ pp.\ \bibinfo {pages}
  {1--52}%
  \bibAnnoteFile{NoStop}{Lockwood:2004}%
\bibitem{Loss:1998}%
  \BibitemOpen
  \bibfield{author}{%
  \bibinfo {author} {\bibfnamefont{D.}~\bibnamefont{Loss}}\ and\ \bibinfo
  {author} {\bibfnamefont{D.~P.}\ \bibnamefont{DiVincenzo}},\ }%
  \bibfield{journal}{%
  \bibinfo {journal} {Phys. Rev. A.}\ }%
  \textbf{\bibinfo {volume} {57}},\ \bibinfo {pages} {120} (\bibinfo {year}
  {1998})%
  \bibAnnoteFile{NoStop}{Loss:1998}%
\bibitem{Yoffe:2002}%
  \BibitemOpen
  \bibfield{author}{%
  \bibinfo {author} {\bibfnamefont{A.~D.}\ \bibnamefont{Yoffe}},\ }%
  \bibfield{journal}{%
  \bibinfo {journal} {Advances in Physics}\ }%
  \textbf{\bibinfo {volume} {51}},\ \bibinfo {pages} {799} (\bibinfo {year}
  {2002})%
  \bibAnnoteFile{NoStop}{Yoffe:2002}%
\bibitem{Frensley:1977}%
  \BibitemOpen
  \bibfield{author}{%
  \bibinfo {author} {\bibfnamefont{W.~R.}\ \bibnamefont{Frensley}}\ and\
  \bibinfo {author} {\bibfnamefont{H.}~\bibnamefont{Kroemer}},\ }%
  \bibfield{journal}{%
  \bibinfo {journal} {Phys. Rev. B.}\ }%
  \textbf{\bibinfo {volume} {16}},\ \bibinfo {pages} {2642} (\bibinfo {year}
  {1977})%
  \bibAnnoteFile{NoStop}{Frensley:1977}%
\bibitem{Bir:1974}%
  \BibitemOpen
  \bibfield{author}{%
  \bibinfo {author} {\bibfnamefont{G.~L.}\ \bibnamefont{Bir}}\ and\ \bibinfo
  {author} {\bibfnamefont{G.~E.}\ \bibnamefont{Pikus}},\ }%
  \emph{\bibinfo {title} {Symmetry and Strain-Induced Effects in
  Semiconductors}}\ (\bibinfo {publisher} {Wiley},\ \bibinfo {address} {New
  York},\ \bibinfo {year} {1976})%
  \bibAnnoteFile{NoStop}{Bir:1974}%
\bibitem{Valentin:2007}%
  \BibitemOpen
  \bibfield{author}{%
  \bibinfo {author} {\bibfnamefont{A.}~\bibnamefont{Valentin}}, \bibinfo
  {author} {\bibfnamefont{J.}~\bibnamefont{S\'{e}e}}, \bibinfo {author}
  {\bibfnamefont{S.}~\bibnamefont{Galdin-Retailleau}},\ and\ \bibinfo {author}
  {\bibfnamefont{P.}~\bibnamefont{Dollfus}},\ }%
  \bibfield{journal}{%
  \bibinfo {journal} {J. Phys.: Conf. Ser.}\ }%
  \textbf{\bibinfo {volume} {92}},\ \bibinfo {pages} {012048} (\bibinfo {year}
  {2007})%
  \bibAnnoteFile{NoStop}{Valentin:2007}%
\bibitem{VanDao:2005}%
  \BibitemOpen
  \bibfield{author}{%
  \bibinfo {author} {\bibfnamefont{L.~V.}\ \bibnamefont{Dao}}, \bibinfo
  {author} {\bibfnamefont{X.}~\bibnamefont{Wen}}, \bibinfo {author}
  {\bibfnamefont{M.~T.~T.}\ \bibnamefont{Do}}, \bibinfo {author}
  {\bibfnamefont{P.}~\bibnamefont{Hannaford}}, \bibinfo {author}
  {\bibfnamefont{E.~C.}\ \bibnamefont{Cho}}, \bibinfo {author}
  {\bibfnamefont{Y.~H.}\ \bibnamefont{Cho}},\ and\ \bibinfo {author}
  {\bibfnamefont{Y.}~\bibnamefont{Huang}},\ }%
  \bibfield{journal}{%
  \bibinfo {journal} {J. Appl. Phys.}\ }%
  \textbf{\bibinfo {volume} {97}},\ \bibinfo {pages} {013501} (\bibinfo {year}
  {2005})%
  \bibAnnoteFile{NoStop}{VanDao:2005}%
\bibitem{Heiss:2005}%
  \BibitemOpen
  \bibfield{author}{%
  \bibinfo {author} {\bibfnamefont{W.~D.}\ \bibnamefont{Heiss}},\ }%
  \emph{\bibinfo {title} {Quantum Dots: a Doorway to Nanoscale Physics}}\
  (\bibinfo {publisher} {Springer},\ \bibinfo {address} {Berlin},\ \bibinfo
  {year} {2005})%
  \bibAnnoteFile{NoStop}{Heiss:2005}%
\bibitem{Zunger:2001}%
  \BibitemOpen
  \bibfield{author}{%
  \bibinfo {author} {\bibfnamefont{A.}~\bibnamefont{Zunger}},\ }%
  \bibfield{journal}{%
  \bibinfo {journal} {Phys. Stat. Sol. (b)}\ }%
  \textbf{\bibinfo {volume} {224}},\ \bibinfo {pages} {727} (\bibinfo {year}
  {2001})%
  \bibAnnoteFile{NoStop}{Zunger:2001}%
\bibitem{Tran:1990}%
  \BibitemOpen
  \bibfield{author}{%
  \bibinfo {author} {\bibfnamefont{D.~B.~T.}\ \bibnamefont{Thoai}}, \bibinfo
  {author} {\bibfnamefont{Y.~Z.}\ \bibnamefont{Hu}},\ and\ \bibinfo {author}
  {\bibfnamefont{S.~W.}\ \bibnamefont{Koch}},\ }%
  \bibfield{journal}{%
  \bibinfo {journal} {Phys. Rev. B.}\ }%
  \textbf{\bibinfo {volume} {42}},\ \bibinfo {pages} {11261} (\bibinfo {year}
  {1990})%
  \bibAnnoteFile{NoStop}{Tran:1990}%
\bibitem{Tit:2010}%
  \BibitemOpen
  \bibfield{author}{%
  \bibinfo {author} {\bibfnamefont{N.}~\bibnamefont{Tit}}, \bibinfo {author}
  {\bibfnamefont{Z.~H.}\ \bibnamefont{Yamani}}, \bibinfo {author}
  {\bibfnamefont{J.}~\bibnamefont{Graham}},\ and\ \bibinfo {author}
  {\bibfnamefont{A.}~\bibnamefont{Ayesh}},\ }%
  \bibfield{journal}{%
  \bibinfo {journal} {Mater. Chem. Phys.}\ }%
  \textbf{\bibinfo {volume} {24}},\ \bibinfo {pages} {927} (\bibinfo {year}
  {2010})%
  \bibAnnoteFile{NoStop}{Tit:2010}%
\bibitem{Buuren:1998}%
  \BibitemOpen
  \bibfield{author}{%
  \bibinfo {author} {\bibfnamefont{T.}~\bibnamefont{van Buuren}}, \bibinfo
  {author} {\bibfnamefont{L.~N.}\ \bibnamefont{Dinh}}, \bibinfo {author}
  {\bibfnamefont{L.~L.}\ \bibnamefont{Chase}}, \bibinfo {author}
  {\bibfnamefont{W.~J.}\ \bibnamefont{Siekhaus}},\ and\ \bibinfo {author}
  {\bibfnamefont{L.~J.}\ \bibnamefont{Terminello}},\ }%
  \bibfield{journal}{%
  \bibinfo {journal} {Phys. Rev. Lett.}\ }%
  \textbf{\bibinfo {volume} {80}},\ \bibinfo {pages} {3803} (\bibinfo {year}
  {1998})%
  \bibAnnoteFile{NoStop}{Buuren:1998}%
\bibitem{Kovalev:1998}%
  \BibitemOpen
  \bibfield{author}{%
  \bibinfo {author} {\bibfnamefont{D.}~\bibnamefont{Kovalev}}, \bibinfo
  {author} {\bibfnamefont{H.}~\bibnamefont{Heckler}}, \bibinfo {author}
  {\bibfnamefont{M.}~\bibnamefont{Ben-Chorin}}, \bibinfo {author}
  {\bibfnamefont{G.}~\bibnamefont{Polisski}}, \bibinfo {author}
  {\bibfnamefont{M.}~\bibnamefont{Schwartzkopff}},\ and\ \bibinfo {author}
  {\bibfnamefont{F.}~\bibnamefont{Koch}},\ }%
  \bibfield{journal}{%
  \bibinfo {journal} {Phys. Rev. Lett.}\ }%
  \textbf{\bibinfo {volume} {81}},\ \bibinfo {pages} {2803} (\bibinfo {year}
  {1998})%
  \bibAnnoteFile{NoStop}{Kovalev:1998}%
\bibitem{Hybertsen:1994}%
  \BibitemOpen
  \bibfield{author}{%
  \bibinfo {author} {\bibfnamefont{M.~S.}\ \bibnamefont{Hybertsen}},\ }%
  \bibfield{journal}{%
  \bibinfo {journal} {Phys. Rev. Lett.}\ }%
  \textbf{\bibinfo {volume} {72}},\ \bibinfo {pages} {1514} (\bibinfo {year}
  {1994})%
  \bibAnnoteFile{NoStop}{Hybertsen:1994}%
\bibitem{Takagahara:1992}%
  \BibitemOpen
  \bibfield{author}{%
  \bibinfo {author} {\bibfnamefont{T.}~\bibnamefont{Takagahara}}\ and\ \bibinfo
  {author} {\bibfnamefont{K.}~\bibnamefont{Takeda}},\ }%
  \bibfield{journal}{%
  \bibinfo {journal} {Phys. Rev. B.}\ }%
  \textbf{\bibinfo {volume} {46}},\ \bibinfo {pages} {15578} (\bibinfo {year}
  {1992})%
  \bibAnnoteFile{NoStop}{Takagahara:1992}%
\bibitem{Barbagiovanni:2011}%
  \BibitemOpen
  \bibfield{author}{%
  \bibinfo {author} {\bibfnamefont{E.~G.}\ \bibnamefont{Barbagiovanni}},
  \bibinfo {author} {\bibfnamefont{L.~V.}\ \bibnamefont{Goncharova}},\ and\
  \bibinfo {author} {\bibfnamefont{P.~J.}\ \bibnamefont{Simpson}},\ }%
  \bibfield{journal}{%
  \bibinfo {journal} {Phys. Rev. B.}\ }%
  \textbf{\bibinfo {volume} {83}},\ \bibinfo {pages} {035112} (\bibinfo {year}
  {2011})%
  \bibAnnoteFile{NoStop}{Barbagiovanni:2011}%
\bibitem{Yu:2001}%
  \BibitemOpen
  \bibfield{author}{%
  \bibinfo {author} {\bibfnamefont{P.~Y.}\ \bibnamefont{Yu}}\ and\ \bibinfo
  {author} {\bibfnamefont{M.}~\bibnamefont{Cardona}},\ }%
  \emph{\bibinfo {title} {Fundamentals of Semiconductors: Physical and Material
  Properties (3rd ed.)}}\ (\bibinfo {publisher} {Springer},\ \bibinfo {address}
  {Berlin},\ \bibinfo {year} {2001})%
  \bibAnnoteFile{NoStop}{Yu:2001}%
\bibitem{Garrido:2004}%
  \BibitemOpen
  \bibfield{author}{%
  \bibinfo {author} {\bibfnamefont{B.}~\bibnamefont{Garrido}}, \bibinfo
  {author} {\bibfnamefont{M.}~\bibnamefont{L\'{o}pez}}, \bibinfo {author}
  {\bibfnamefont{A.}~\bibnamefont{P\'{e}rez-Rodr\'{i}guez}}, \bibinfo {author}
  {\bibfnamefont{C.}~\bibnamefont{Garc\'{i}a}}, \bibinfo {author}
  {\bibfnamefont{P.}~\bibnamefont{Pellegrino}}, \bibinfo {author}
  {\bibfnamefont{R.}~\bibnamefont{Ferr\`{e}}}, \bibinfo {author}
  {\bibfnamefont{J.~A.}\ \bibnamefont{Moreno}}, \bibinfo {author}
  {\bibfnamefont{J.~R.}\ \bibnamefont{Morante}}, \bibinfo {author}
  {\bibfnamefont{C.}~\bibnamefont{Bonafos}}, \bibinfo {author}
  {\bibfnamefont{M.}~\bibnamefont{Carrada}}, \bibinfo {author}
  {\bibfnamefont{A.}~\bibnamefont{Claverie}}, \bibinfo {author}
  {\bibfnamefont{J.}~\bibnamefont{de~la Torre}},\ and\ \bibinfo {author}
  {\bibfnamefont{A.}~\bibnamefont{Souifi}},\ }%
  \bibfield{journal}{%
  \bibinfo {journal} {Nuc. Instr. and Meth. in Phys. Res. B}\ }%
  \textbf{\bibinfo {volume} {216}},\ \bibinfo {pages} {213} (\bibinfo {year}
  {2004})%
  \bibAnnoteFile{NoStop}{Garrido:2004}%
\bibitem{Takagi:1990}%
  \BibitemOpen
  \bibfield{author}{%
  \bibinfo {author} {\bibfnamefont{H.}~\bibnamefont{Takagi}}, \bibinfo {author}
  {\bibfnamefont{H.}~\bibnamefont{Ogawa}}, \bibinfo {author}
  {\bibfnamefont{Y.}~\bibnamefont{Yamazaki}}, \bibinfo {author}
  {\bibfnamefont{A.}~\bibnamefont{Ishizaki}},\ and\ \bibinfo {author}
  {\bibfnamefont{T.}~\bibnamefont{Nakagiri}},\ }%
  \bibfield{journal}{%
  \bibinfo {journal} {Appl. Phys. Lett.}\ }%
  \textbf{\bibinfo {volume} {56}},\ \bibinfo {pages} {2379} (\bibinfo {year}
  {1990})%
  \bibAnnoteFile{NoStop}{Takagi:1990}%
\bibitem{Lu:2002}%
  \BibitemOpen
  \bibfield{author}{%
  \bibinfo {author} {\bibfnamefont{Z.~H.}\ \bibnamefont{Lu}}\ and\ \bibinfo
  {author} {\bibfnamefont{D.}~\bibnamefont{Grozea}},\ }%
  \bibfield{journal}{%
  \bibinfo {journal} {Appl. Phys. Lett.}\ }%
  \textbf{\bibinfo {volume} {80}},\ \bibinfo {pages} {255} (\bibinfo {year}
  {2002})%
  \bibAnnoteFile{NoStop}{Lu:2002}%
\bibitem{Barbagiovanni:2011:1}%
  \BibitemOpen
  \bibfield{author}{%
  \bibinfo {author} {\bibfnamefont{E.}~\bibnamefont{Barbagiovanni}}, \bibinfo
  {author} {\bibfnamefont{S.}~\bibnamefont{Dedyulin}}, \bibinfo {author}
  {\bibfnamefont{P.}~\bibnamefont{Simpson}},\ and\ \bibinfo {author}
  {\bibfnamefont{L.}~\bibnamefont{Goncharova}},\ }%
  \bibfield{journal}{%
  \bibinfo {journal} {Nuc. Instr. and Meth. in Phys. Res. B},\ \bibinfo {pages}
  {doi:10.1016/j.nimb.2011.01.036}}%
   (\bibinfo {year} {2011})%
  \bibAnnoteFile{NoStop}{Barbagiovanni:2011:1}%
\bibitem{Wang:1994:1}%
  \BibitemOpen
  \bibfield{author}{%
  \bibinfo {author} {\bibfnamefont{Q.}~\bibnamefont{Wang}}, \bibinfo {author}
  {\bibfnamefont{F.}~\bibnamefont{Lu}}, \bibinfo {author}
  {\bibfnamefont{D.}~\bibnamefont{Gong}}, \bibinfo {author}
  {\bibfnamefont{X.}~\bibnamefont{Chen}}, \bibinfo {author}
  {\bibfnamefont{J.}~\bibnamefont{Wang}}, \bibinfo {author}
  {\bibfnamefont{H.}~\bibnamefont{Sun}},\ and\ \bibinfo {author}
  {\bibfnamefont{X.}~\bibnamefont{Wang}},\ }%
  \bibfield{journal}{%
  \bibinfo {journal} {Phys. Rev. B.}\ }%
  \textbf{\bibinfo {volume} {50}},\ \bibinfo {pages} {18226} (\bibinfo {year}
  {1994})%
  \bibAnnoteFile{NoStop}{Wang:1994:1}%
\bibitem{Jin:1999}%
  \BibitemOpen
  \bibfield{author}{%
  \bibinfo {author} {\bibfnamefont{G.}~\bibnamefont{Jin}}, \bibinfo {author}
  {\bibfnamefont{Y.~S.}\ \bibnamefont{Tang}}, \bibinfo {author}
  {\bibfnamefont{J.~L.}\ \bibnamefont{Liu}},\ and\ \bibinfo {author}
  {\bibfnamefont{K.~L.}\ \bibnamefont{Wang}},\ }%
  \bibfield{journal}{%
  \bibinfo {journal} {Appl. Phys. Lett.}\ }%
  \textbf{\bibinfo {volume} {74}},\ \bibinfo {pages} {2471} (\bibinfo {year}
  {1999})%
  \bibAnnoteFile{NoStop}{Jin:1999}%
\bibitem{Min:1996:1}%
  \BibitemOpen
  \bibfield{author}{%
  \bibinfo {author} {\bibfnamefont{K.~S.}\ \bibnamefont{Min}}, \bibinfo
  {author} {\bibfnamefont{K.~V.}\ \bibnamefont{Shcheglov}}, \bibinfo {author}
  {\bibfnamefont{C.~M.}\ \bibnamefont{Yang}}, \bibinfo {author}
  {\bibfnamefont{H.~A.}\ \bibnamefont{Atwater}}, \bibinfo {author}
  {\bibfnamefont{M.~L.}\ \bibnamefont{Brongersma}},\ and\ \bibinfo {author}
  {\bibfnamefont{A.}~\bibnamefont{Polman}},\ }%
  \bibfield{journal}{%
  \bibinfo {journal} {Appl. Phys. Lett.}\ }%
  \textbf{\bibinfo {volume} {68}},\ \bibinfo {pages} {2511} (\bibinfo {year}
  {1996})%
  \bibAnnoteFile{NoStop}{Min:1996:1}%
\bibitem{Rowell:2009}%
  \BibitemOpen
  \bibfield{author}{%
  \bibinfo {author} {\bibfnamefont{N.~L.}\ \bibnamefont{Rowell}}, \bibinfo
  {author} {\bibfnamefont{D.~J.}\ \bibnamefont{Lockwood}}, \bibinfo {author}
  {\bibfnamefont{I.}~\bibnamefont{Berbezier}}, \bibinfo {author}
  {\bibfnamefont{P.~D.}\ \bibnamefont{Szkutnik}},\ and\ \bibinfo {author}
  {\bibfnamefont{A.}~\bibnamefont{Ronda}},\ }%
  \bibfield{journal}{%
  \bibinfo {journal} {J. Electrochem. Soc.}\ }%
  \textbf{\bibinfo {volume} {156}},\ \bibinfo {pages} {H913} (\bibinfo {year}
  {2009})%
  \bibAnnoteFile{NoStop}{Rowell:2009}%
\bibitem{Lu:1995}%
  \BibitemOpen
  \bibfield{author}{%
  \bibinfo {author} {\bibfnamefont{Z.~H.}\ \bibnamefont{Lu}}, \bibinfo {author}
  {\bibfnamefont{D.~J.}\ \bibnamefont{Lockwood}},\ and\ \bibinfo {author}
  {\bibfnamefont{J.~M.}\ \bibnamefont{Baribeau}},\ }%
  \bibfield{journal}{%
  \bibinfo {journal} {Nature}\ }%
  \textbf{\bibinfo {volume} {378}},\ \bibinfo {pages} {258} (\bibinfo {year}
  {1995})%
  \bibAnnoteFile{NoStop}{Lu:1995}%
\bibitem{Lockwood:1996}%
  \BibitemOpen
  \bibfield{author}{%
  \bibinfo {author} {\bibfnamefont{D.~J.}\ \bibnamefont{Lockwood}}, \bibinfo
  {author} {\bibfnamefont{Z.~H.}\ \bibnamefont{Lu}},\ and\ \bibinfo {author}
  {\bibfnamefont{J.~M.}\ \bibnamefont{Baribeau}},\ }%
  \bibfield{journal}{%
  \bibinfo {journal} {Phys. Rev. Lett.}\ }%
  \textbf{\bibinfo {volume} {76}},\ \bibinfo {pages} {539} (\bibinfo {year}
  {1996})%
  \bibAnnoteFile{NoStop}{Lockwood:1996}%
\bibitem{Lockwood:2003}%
  \BibitemOpen
  \bibfield{author}{%
  \bibinfo {author} {\bibfnamefont{D.~J.}\ \bibnamefont{Lockwood}}, \bibinfo
  {author} {\bibfnamefont{M.~W.~C.}\ \bibnamefont{Dharma-wardana}}, \bibinfo
  {author} {\bibfnamefont{Z.~H.}\ \bibnamefont{Lu}}, \bibinfo {author}
  {\bibfnamefont{D.~H.}\ \bibnamefont{Grozea}}, \bibinfo {author}
  {\bibfnamefont{P.}~\bibnamefont{Carrier}},\ and\ \bibinfo {author}
  {\bibfnamefont{L.~J.}\ \bibnamefont{Lewis}},\ }%
  \bibfield{journal}{%
  \bibinfo {journal} {Mater. Res. Soc. Symp. Proc.}\ }%
  \textbf{\bibinfo {volume} {737}},\ \bibinfo {pages} {F1.1.1} (\bibinfo {year}
  {2003})%
  \bibAnnoteFile{NoStop}{Lockwood:2003}%
\bibitem{Mokry:2009}%
  \BibitemOpen
  \bibfield{author}{%
  \bibinfo {author} {\bibfnamefont{C.~R.}\ \bibnamefont{Mokry}}, \bibinfo
  {author} {\bibfnamefont{P.~J.}\ \bibnamefont{Simpson}},\ and\ \bibinfo
  {author} {\bibfnamefont{A.~P.}\ \bibnamefont{Knights}},\ }%
  \bibfield{journal}{%
  \bibinfo {journal} {J. Appl. Phys.}\ }%
  \textbf{\bibinfo {volume} {105}},\ \bibinfo {pages} {114301.1} (\bibinfo
  {year} {2009})%
  \bibAnnoteFile{NoStop}{Mokry:2009}%
\bibitem{Linnros:1999}%
  \BibitemOpen
  \bibfield{author}{%
  \bibinfo {author} {\bibfnamefont{J.}~\bibnamefont{Linnros}}, \bibinfo
  {author} {\bibfnamefont{N.}~\bibnamefont{Lalic}}, \bibinfo {author}
  {\bibfnamefont{A.}~\bibnamefont{Galeckas}},\ and\ \bibinfo {author}
  {\bibfnamefont{V.}~\bibnamefont{Grivickas}},\ }%
  \bibfield{journal}{%
  \bibinfo {journal} {J. Appl. Phys.}\ }%
  \textbf{\bibinfo {volume} {86}},\ \bibinfo {pages} {6128} (\bibinfo {year}
  {1999})%
  \bibAnnoteFile{NoStop}{Linnros:1999}%
\bibitem{Kamenev:2004}%
  \BibitemOpen
  \bibfield{author}{%
  \bibinfo {author} {\bibfnamefont{B.~V.}\ \bibnamefont{Kamenev}}, \bibinfo
  {author} {\bibfnamefont{G.~F.}\ \bibnamefont{Grom}}, \bibinfo {author}
  {\bibfnamefont{D.~J.}\ \bibnamefont{Lockwood}}, \bibinfo {author}
  {\bibfnamefont{J.~P.}\ \bibnamefont{McCafrey}}, \bibinfo {author}
  {\bibfnamefont{B.}~\bibnamefont{Laikhtman}},\ and\ \bibinfo {author}
  {\bibfnamefont{L.}~\bibnamefont{Tsybeskov}},\ }%
  \bibfield{journal}{%
  \bibinfo {journal} {Phys. Rev. B.}\ }%
  \textbf{\bibinfo {volume} {69}},\ \bibinfo {pages} {235306} (\bibinfo {year}
  {2004})%
  \bibAnnoteFile{NoStop}{Kamenev:2004}%
\bibitem{Iacona:2000}%
  \BibitemOpen
  \bibfield{author}{%
  \bibinfo {author} {\bibfnamefont{F.}~\bibnamefont{Iacona}}, \bibinfo {author}
  {\bibfnamefont{G.}~\bibnamefont{Franzo}},\ and\ \bibinfo {author}
  {\bibfnamefont{C.}~\bibnamefont{Spinella}},\ }%
  \bibfield{journal}{%
  \bibinfo {journal} {J. Appl. Phys.}\ }%
  \textbf{\bibinfo {volume} {87}},\ \bibinfo {pages} {1295} (\bibinfo {year}
  {2000})%
  \bibAnnoteFile{NoStop}{Iacona:2000}%
\bibitem{Park:2001}%
  \BibitemOpen
  \bibfield{author}{%
  \bibinfo {author} {\bibfnamefont{N.~M.}\ \bibnamefont{Park}}, \bibinfo
  {author} {\bibfnamefont{C.~J.}\ \bibnamefont{Choi}}, \bibinfo {author}
  {\bibfnamefont{T.~Y.}\ \bibnamefont{Seong}},\ and\ \bibinfo {author}
  {\bibfnamefont{S.~J.}\ \bibnamefont{Park}},\ }%
  \bibfield{journal}{%
  \bibinfo {journal} {Phys. Rev. Lett.}\ }%
  \textbf{\bibinfo {volume} {86}},\ \bibinfo {pages} {1355} (\bibinfo {year}
  {2001})%
  \bibAnnoteFile{NoStop}{Park:2001}%
\bibitem{Seo:2009}%
  \BibitemOpen
  \bibfield{author}{%
  \bibinfo {author} {\bibfnamefont{H.~S.}\ \bibnamefont{Seo}}, \bibinfo
  {author} {\bibfnamefont{X.}~\bibnamefont{Li}}, \bibinfo {author}
  {\bibfnamefont{H.~D.}\ \bibnamefont{Um}}, \bibinfo {author}
  {\bibfnamefont{B.}~\bibnamefont{Yoo}}, \bibinfo {author} {\bibfnamefont{J.~H.
  K. K.~P.}\ \bibnamefont{Kim}}, \bibinfo {author} {\bibfnamefont{Y.~W.}\
  \bibnamefont{Cho}},\ and\ \bibinfo {author} {\bibfnamefont{J.~H.}\
  \bibnamefont{Lee}},\ }%
  \bibfield{journal}{%
  \bibinfo {journal} {Mater. Lett.}\ }%
  \textbf{\bibinfo {volume} {63}},\ \bibinfo {pages} {2567} (\bibinfo {year}
  {2009})%
  \bibAnnoteFile{NoStop}{Seo:2009}%
\bibitem{Hamilton:1995}%
  \BibitemOpen
  \bibfield{author}{%
  \bibinfo {author} {\bibfnamefont{B.}~\bibnamefont{Hamilton}},\ }%
  \bibfield{journal}{%
  \bibinfo {journal} {Semicond. Sci. Technol.}\ }%
  \textbf{\bibinfo {volume} {10}},\ \bibinfo {pages} {1187} (\bibinfo {year}
  {1995})%
  \bibAnnoteFile{NoStop}{Hamilton:1995}%
\bibitem{Lockwood:1995}%
  \BibitemOpen
  \bibfield{author}{%
  \bibinfo {author} {\bibfnamefont{D.~J.}\ \bibnamefont{Lockwood}}\ and\
  \bibinfo {author} {\bibfnamefont{A.~G.}\ \bibnamefont{Wang}},\ }%
  \bibfield{journal}{%
  \bibinfo {journal} {Solid State Commun.}\ }%
  \textbf{\bibinfo {volume} {94}},\ \bibinfo {pages} {905} (\bibinfo {year}
  {1995})%
  \bibAnnoteFile{NoStop}{Lockwood:1995}%
\bibitem{Schuppler:1994}%
  \BibitemOpen
  \bibfield{author}{%
  \bibinfo {author} {\bibfnamefont{S.}~\bibnamefont{Schuppler}}, \bibinfo
  {author} {\bibfnamefont{S.~L.}\ \bibnamefont{Friedman}}, \bibinfo {author}
  {\bibfnamefont{M.~A.}\ \bibnamefont{Marcus}}, \bibinfo {author}
  {\bibfnamefont{D.~L.}\ \bibnamefont{Adler}}, \bibinfo {author}
  {\bibfnamefont{Y.-H.}\ \bibnamefont{Xie}}, \bibinfo {author}
  {\bibfnamefont{F.~M.}\ \bibnamefont{Ross}}, \bibinfo {author}
  {\bibfnamefont{T.~D.}\ \bibnamefont{Harris}}, \bibinfo {author}
  {\bibfnamefont{W.~L.}\ \bibnamefont{Brown}}, \bibinfo {author}
  {\bibfnamefont{Y.~J.}\ \bibnamefont{Chabal}}, \bibinfo {author}
  {\bibfnamefont{L.~E.}\ \bibnamefont{Brus}},\ and\ \bibinfo {author}
  {\bibfnamefont{P.~H.}\ \bibnamefont{Citrin}},\ }%
  \bibfield{journal}{%
  \bibinfo {journal} {Phys. Rev. Lett.}\ }%
  \textbf{\bibinfo {volume} {72}},\ \bibinfo {pages} {2648} (\bibinfo {year}
  {1994})%
  \bibAnnoteFile{NoStop}{Schuppler:1994}%
\bibitem{Ma:2003}%
  \BibitemOpen
  \bibfield{author}{%
  \bibinfo {author} {\bibfnamefont{D.~D.~D.}\ \bibnamefont{Ma}}, \bibinfo
  {author} {\bibfnamefont{C.~S.}\ \bibnamefont{Lee}}, \bibinfo {author}
  {\bibfnamefont{F.~C.~K.}\ \bibnamefont{Au}}, \bibinfo {author}
  {\bibfnamefont{S.~Y.}\ \bibnamefont{Tong}},\ and\ \bibinfo {author}
  {\bibfnamefont{S.~T.}\ \bibnamefont{Lee}},\ }%
  \bibfield{journal}{%
  \bibinfo {journal} {Science}\ }%
  \textbf{\bibinfo {volume} {299}},\ \bibinfo {pages} {1874} (\bibinfo {year}
  {2003})%
  \bibAnnoteFile{NoStop}{Ma:2003}%
\bibitem{Yoshioka:2011}%
  \BibitemOpen
  \bibfield{author}{%
  \bibinfo {author} {\bibfnamefont{H.}~\bibnamefont{Yoshioka}}, \bibinfo
  {author} {\bibfnamefont{N.}~\bibnamefont{Morioka}}, \bibinfo {author}
  {\bibfnamefont{J.}~\bibnamefont{Suda}},\ and\ \bibinfo {author}
  {\bibfnamefont{T.}~\bibnamefont{Kimoto}},\ }%
  \bibfield{journal}{%
  \bibinfo {journal} {J. Appl. Phys.}\ }%
  \textbf{\bibinfo {volume} {109}},\ \bibinfo {pages} {064312} (\bibinfo {year}
  {2011})%
  \bibAnnoteFile{NoStop}{Yoshioka:2011}%
\bibitem{Lockwood:1994}%
  \BibitemOpen
  \bibfield{author}{%
  \bibinfo {author} {\bibfnamefont{D.~J.}\ \bibnamefont{Lockwood}},\ }%
  \bibfield{journal}{%
  \bibinfo {journal} {Solid State Commun.}\ }%
  \textbf{\bibinfo {volume} {92}},\ \bibinfo {pages} {101} (\bibinfo {year}
  {1994})%
  \bibAnnoteFile{NoStop}{Lockwood:1994}%
\bibitem{Takeoka:1998}%
  \BibitemOpen
  \bibfield{author}{%
  \bibinfo {author} {\bibfnamefont{S.}~\bibnamefont{Takeoka}}, \bibinfo
  {author} {\bibfnamefont{M.}~\bibnamefont{Fujii}}, \bibinfo {author}
  {\bibfnamefont{S.}~\bibnamefont{Hayashi}},\ and\ \bibinfo {author}
  {\bibfnamefont{K.}~\bibnamefont{Yamamoto}},\ }%
  \bibfield{journal}{%
  \bibinfo {journal} {Phys. Rev. B.}\ }%
  \textbf{\bibinfo {volume} {58}},\ \bibinfo {pages} {7921} (\bibinfo {year}
  {1998})%
  \bibAnnoteFile{NoStop}{Takeoka:1998}%
\bibitem{Bostedt:2004}%
  \BibitemOpen
  \bibfield{author}{%
  \bibinfo {author} {\bibfnamefont{C.}~\bibnamefont{Bostedt}}, \bibinfo
  {author} {\bibfnamefont{T.}~\bibnamefont{van Buuren}}, \bibinfo {author}
  {\bibfnamefont{T.~M.}\ \bibnamefont{Willey}}, \bibinfo {author}
  {\bibfnamefont{N.}~\bibnamefont{Franco}}, \bibinfo {author}
  {\bibfnamefont{L.~J.}\ \bibnamefont{Terminello}}, \bibinfo {author}
  {\bibfnamefont{C.}~\bibnamefont{Heske}},\ and\ \bibinfo {author}
  {\bibfnamefont{T.}~\bibnamefont{M\"{o}ller}},\ }%
  \bibfield{journal}{%
  \bibinfo {journal} {Appl. Phys. Lett.}\ }%
  \textbf{\bibinfo {volume} {84}},\ \bibinfo {pages} {4056} (\bibinfo {year}
  {2004})%
  \bibAnnoteFile{NoStop}{Bostedt:2004}%
\bibitem{Eaton:2010}%
  \BibitemOpen
  \bibfield{author}{%
  \bibinfo {author} {\bibfnamefont{P.}~\bibnamefont{Eaton}}\ and\ \bibinfo
  {author} {\bibfnamefont{P.}~\bibnamefont{West}},\ }%
  \emph{\bibinfo {title} {Atomic Force Microscopy}}\ (\bibinfo {publisher}
  {Oxford University Press},\ \bibinfo {address} {Oxford},\ \bibinfo {year}
  {2010})\ p.\ \bibinfo {pages} {110}%
  \bibAnnoteFile{NoStop}{Eaton:2010}%
\bibitem{Street:1991}%
  \BibitemOpen
  \bibfield{author}{%
  \bibinfo {author} {\bibfnamefont{R.~A.}\ \bibnamefont{Street}},\ }%
  \emph{\bibinfo {title} {Hydrogenated Amorphous Silicon}}\ (\bibinfo
  {publisher} {Cambridge University Press},\ \bibinfo {address} {Cambridge},\
  \bibinfo {year} {1991})%
  \bibAnnoteFile{NoStop}{Street:1991}%
\bibitem{See:2002}%
  \BibitemOpen
  \bibfield{author}{%
  \bibinfo {author} {\bibfnamefont{J.}~\bibnamefont{S\'{e}e}}, \bibinfo
  {author} {\bibfnamefont{P.}~\bibnamefont{Dollfus}},\ and\ \bibinfo {author}
  {\bibfnamefont{S.}~\bibnamefont{Galdin}},\ }%
  \bibfield{journal}{%
  \bibinfo {journal} {Phys. Rev. B.}\ }%
  \textbf{\bibinfo {volume} {66}},\ \bibinfo {pages} {193307} (\bibinfo {year}
  {2002})%
  \bibAnnoteFile{NoStop}{See:2002}%
\bibitem{Lassen:2009}%
  \BibitemOpen
  \bibfield{author}{%
  \bibinfo {author} {\bibfnamefont{B.}~\bibnamefont{Lassen}}, \bibinfo {author}
  {\bibfnamefont{R.~V.~N.}\ \bibnamefont{Melnik}},\ and\ \bibinfo {author}
  {\bibfnamefont{M.}~\bibnamefont{Willatzen}},\ }%
  \bibfield{journal}{%
  \bibinfo {journal} {Commun. Comput. Phys.}\ }%
  \textbf{\bibinfo {volume} {6}},\ \bibinfo {pages} {699} (\bibinfo {year}
  {2009})%
  \bibAnnoteFile{NoStop}{Lassen:2009}%
\bibitem{Delerue:1993}%
  \BibitemOpen
  \bibfield{author}{%
  \bibinfo {author} {\bibfnamefont{C.}~\bibnamefont{Delerue}}, \bibinfo
  {author} {\bibfnamefont{G.}~\bibnamefont{Allan}},\ and\ \bibinfo {author}
  {\bibfnamefont{M.}~\bibnamefont{Lannoo}},\ }%
  \bibfield{journal}{%
  \bibinfo {journal} {Phys. Rev. B.}\ }%
  \textbf{\bibinfo {volume} {48}},\ \bibinfo {pages} {11024} (\bibinfo {year}
  {1993})%
  \bibAnnoteFile{NoStop}{Delerue:1993}%
\bibitem{Tomic:2011}%
  \BibitemOpen
  \bibfield{author}{%
  \bibinfo {author} {\bibfnamefont{S.}~\bibnamefont{Tomi\'{c}}}\ and\ \bibinfo
  {author} {\bibfnamefont{N.}~\bibnamefont{Vukmirovi\'{c}}},\ }%
  \bibfield{journal}{%
  \bibinfo {journal} {J. Appl. Phys.}\ }%
  \textbf{\bibinfo {volume} {110}},\ \bibinfo {pages} {053710} (\bibinfo {year}
  {2011})%
  \bibAnnoteFile{NoStop}{Tomic:2011}%
\bibitem{Bulutay:2007}%
  \BibitemOpen
  \bibfield{author}{%
  \bibinfo {author} {\bibfnamefont{C.}~\bibnamefont{Bulutay}},\ }%
  \bibfield{journal}{%
  \bibinfo {journal} {Phys. Rev. B.}\ }%
  \textbf{\bibinfo {volume} {76}},\ \bibinfo {pages} {17} (\bibinfo {year}
  {2007})%
  \bibAnnoteFile{NoStop}{Bulutay:2007}%
\bibitem{Furukawa:1988}%
  \BibitemOpen
  \bibfield{author}{%
  \bibinfo {author} {\bibfnamefont{S.}~\bibnamefont{Furukawa}}\ and\ \bibinfo
  {author} {\bibfnamefont{T.}~\bibnamefont{Miyasato}},\ }%
  \bibfield{journal}{%
  \bibinfo {journal} {Phys. Rev. B.}\ }%
  \textbf{\bibinfo {volume} {38}},\ \bibinfo {pages} {5726} (\bibinfo {year}
  {1988})%
  \bibAnnoteFile{NoStop}{Furukawa:1988}%
\bibitem{Hong:2003}%
  \BibitemOpen
  \bibfield{author}{%
  \bibinfo {author} {\bibfnamefont{C.~C.}\ \bibnamefont{Hong}}, \bibinfo
  {author} {\bibfnamefont{W.~J.}\ \bibnamefont{Liao}},\ and\ \bibinfo {author}
  {\bibfnamefont{J.~G.}\ \bibnamefont{Hwu}},\ }%
  \bibfield{journal}{%
  \bibinfo {journal} {Appl. Phys. Lett.}\ }%
  \textbf{\bibinfo {volume} {82}},\ \bibinfo {pages} {3916} (\bibinfo {year}
  {2003})%
  \bibAnnoteFile{NoStop}{Hong:2003}%
\bibitem{Lopez:2002}%
  \BibitemOpen
  \bibfield{author}{%
  \bibinfo {author} {\bibfnamefont{M.}~\bibnamefont{L\'{o}pez}}, \bibinfo
  {author} {\bibfnamefont{B.}~\bibnamefont{Garrido}}, \bibinfo {author}
  {\bibfnamefont{C.}~\bibnamefont{Garc\'{i}a}}, \bibinfo {author}
  {\bibfnamefont{P.}~\bibnamefont{Pellegrino}}, \bibinfo {author}
  {\bibfnamefont{A.}~\bibnamefont{P\'{e}rez-Rodr\'{i}guez}}, \bibinfo {author}
  {\bibfnamefont{J.~R.}\ \bibnamefont{Morante}}, \bibinfo {author}
  {\bibfnamefont{C.}~\bibnamefont{Bonafos}}, \bibinfo {author}
  {\bibfnamefont{M.}~\bibnamefont{Carrada}},\ and\ \bibinfo {author}
  {\bibfnamefont{A.}~\bibnamefont{Claverie}},\ }%
  \bibfield{journal}{%
  \bibinfo {journal} {Appl. Phys. Lett.}\ }%
  \textbf{\bibinfo {volume} {80}},\ \bibinfo {pages} {1637} (\bibinfo {year}
  {2002})%
  \bibAnnoteFile{NoStop}{Lopez:2002}%
\bibitem{Kun:2005}%
  \BibitemOpen
  \bibfield{author}{%
  \bibinfo {author} {\bibfnamefont{L.~Z.}\ \bibnamefont{Kun}}, \bibinfo
  {author} {\bibfnamefont{K.~Y.}\ \bibnamefont{Lan}}, \bibinfo {author}
  {\bibfnamefont{C.}~\bibnamefont{Hao}}, \bibinfo {author}
  {\bibfnamefont{H.}~\bibnamefont{Ming}},\ and\ \bibinfo {author}
  {\bibfnamefont{Q.}~\bibnamefont{Yu}},\ }%
  \bibfield{journal}{%
  \bibinfo {journal} {Chin. Phys. Lett.}\ }%
  \textbf{\bibinfo {volume} {22}},\ \bibinfo {pages} {984} (\bibinfo {year}
  {2005})%
  \bibAnnoteFile{NoStop}{Kun:2005}%
\bibitem{Wolkin:1999}%
  \BibitemOpen
  \bibfield{author}{%
  \bibinfo {author} {\bibfnamefont{M.~V.}\ \bibnamefont{Wolkin}}, \bibinfo
  {author} {\bibfnamefont{J.}~\bibnamefont{Jorne}}, \bibinfo {author}
  {\bibfnamefont{P.~M.}\ \bibnamefont{Fauchet}}, \bibinfo {author}
  {\bibfnamefont{G.}~\bibnamefont{Allan}},\ and\ \bibinfo {author}
  {\bibfnamefont{C.}~\bibnamefont{Delerue}},\ }%
  \bibfield{journal}{%
  \bibinfo {journal} {Phys. Rev. Lett.}\ }%
  \textbf{\bibinfo {volume} {82}},\ \bibinfo {pages} {197} (\bibinfo {year}
  {1999})%
  \bibAnnoteFile{NoStop}{Wolkin:1999}%
\bibitem{Svechnikov:1998}%
  \BibitemOpen
  \bibfield{author}{%
  \bibinfo {author} {\bibfnamefont{S.~V.}\ \bibnamefont{Svechnikov}}, \bibinfo
  {author} {\bibfnamefont{E.~B.}\ \bibnamefont{Kaganovich}},\ and\ \bibinfo
  {author} {\bibfnamefont{E.~G.}\ \bibnamefont{Manoilov}},\ }%
  \bibfield{journal}{%
  \bibinfo {journal} {Semi. Phy. Quantum Elect. Optoelect.}\ }%
  \textbf{\bibinfo {volume} {1}},\ \bibinfo {pages} {13} (\bibinfo {year}
  {1998})%
  \bibAnnoteFile{NoStop}{Svechnikov:1998}%
\bibitem{Sharp:2005}%
  \BibitemOpen
  \bibfield{author}{%
  \bibinfo {author} {\bibfnamefont{I.~D.}\ \bibnamefont{Sharp}}, \bibinfo
  {author} {\bibfnamefont{D.~O.}\ \bibnamefont{Yi}}, \bibinfo {author}
  {\bibfnamefont{Q.}~\bibnamefont{Xu}}, \bibinfo {author}
  {\bibfnamefont{C.~Y.}\ \bibnamefont{Liao}}, \bibinfo {author}
  {\bibfnamefont{J.~W.}\ \bibnamefont{Beeman}}, \bibinfo {author}
  {\bibfnamefont{Z.}~\bibnamefont{Liliental-Weber}}, \bibinfo {author}
  {\bibfnamefont{K.~M.}\ \bibnamefont{Yu}}, \bibinfo {author}
  {\bibfnamefont{D.~N.}\ \bibnamefont{Zakharov}}, \bibinfo {author}
  {\bibfnamefont{J.~W.}\ \bibnamefont{Ager}}, \bibinfo {author}
  {\bibfnamefont{D.~C.}\ \bibnamefont{Chrzan}},\ and\ \bibinfo {author}
  {\bibfnamefont{E.~E.}\ \bibnamefont{Haller}},\ }%
  \bibfield{journal}{%
  \bibinfo {journal} {Appl. Phys. Lett.}\ }%
  \textbf{\bibinfo {volume} {86}},\ \bibinfo {pages} {063107} (\bibinfo {year}
  {2005})%
  \bibAnnoteFile{NoStop}{Sharp:2005}%
\bibitem{Saar:2005}%
  \BibitemOpen
  \bibfield{author}{%
  \bibinfo {author} {\bibfnamefont{A.}~\bibnamefont{Sa'ar}}, \bibinfo {author}
  {\bibfnamefont{Y.}~\bibnamefont{Reichman}}, \bibinfo {author}
  {\bibfnamefont{M.}~\bibnamefont{Dovrat}}, \bibinfo {author}
  {\bibfnamefont{D.}~\bibnamefont{Krapf}}, \bibinfo {author}
  {\bibfnamefont{J.}~\bibnamefont{Jedrzejewski}},\ and\ \bibinfo {author}
  {\bibfnamefont{I.}~\bibnamefont{Balberg}},\ }%
  \bibfield{journal}{%
  \bibinfo {journal} {Nano Lett.}\ }%
  \textbf{\bibinfo {volume} {5}},\ \bibinfo {pages} {2443} (\bibinfo {year}
  {2005})%
  \bibAnnoteFile{NoStop}{Saar:2005}%
\bibitem{Kanemitsu:2002}%
  \BibitemOpen
  \bibfield{author}{%
  \bibinfo {author} {\bibfnamefont{Y.}~\bibnamefont{Kanemitsu}},\ }%
  \bibfield{journal}{%
  \bibinfo {journal} {J. Luminescence}\ }%
  \textbf{\bibinfo {volume} {100}},\ \bibinfo {pages} {209} (\bibinfo {year}
  {2002})%
  \bibAnnoteFile{NoStop}{Kanemitsu:2002}%
\bibitem{Seino:2011}%
  \BibitemOpen
  \bibfield{author}{%
  \bibinfo {author} {\bibfnamefont{K.}~\bibnamefont{Seino}}\ and\ \bibinfo
  {author} {\bibfnamefont{F.}~\bibnamefont{Bechstedt}},\ }%
  \bibfield{journal}{%
  \bibinfo {journal} {Semicond. Sci. Technol.}\ }%
  \textbf{\bibinfo {volume} {26}},\ \bibinfo {pages} {014024} (\bibinfo {year}
  {2011})%
  \bibAnnoteFile{NoStop}{Seino:2011}%
\bibitem{Rossner:2003}%
  \BibitemOpen
  \bibfield{author}{%
  \bibinfo {author} {\bibfnamefont{B.}~\bibnamefont{R\"{o}$\ss$ner}}, \bibinfo
  {author} {\bibfnamefont{G.}~\bibnamefont{Isella}},\ and\ \bibinfo {author}
  {\bibfnamefont{H.}~\bibnamefont{von K\"{a}nel}},\ }%
  \bibfield{journal}{%
  \bibinfo {journal} {Appl. Phys. Lett.}\ }%
  \textbf{\bibinfo {volume} {82}},\ \bibinfo {pages} {754} (\bibinfo {year}
  {2003})%
  \bibAnnoteFile{NoStop}{Rossner:2003}%
\bibitem{Yang:2004}%
  \BibitemOpen
  \bibfield{author}{%
  \bibinfo {author} {\bibfnamefont{Z.}~\bibnamefont{Yang}}, \bibinfo {author}
  {\bibfnamefont{Y.}~\bibnamefont{Shi}}, \bibinfo {author}
  {\bibfnamefont{J.}~\bibnamefont{Liu}}, \bibinfo {author}
  {\bibfnamefont{B.}~\bibnamefont{Yan}}, \bibinfo {author}
  {\bibfnamefont{R.}~\bibnamefont{Zhang}}, \bibinfo {author}
  {\bibfnamefont{Y.}~\bibnamefont{Zheng}},\ and\ \bibinfo {author}
  {\bibfnamefont{K.}~\bibnamefont{Wang}},\ }%
  \bibfield{journal}{%
  \bibinfo {journal} {Mater. Lett.}\ }%
  \textbf{\bibinfo {volume} {58}},\ \bibinfo {pages} {3765} (\bibinfo {year}
  {2004})%
  \bibAnnoteFile{NoStop}{Yang:2004}%
\bibitem{Ley:1984}%
  \BibitemOpen
  \bibfield{author}{%
  \bibinfo {author} {\bibfnamefont{L.}~\bibnamefont{Ley}},\ }%
  \enquote{\bibinfo {title} {The physics of hydrogenated amorphous silicon
  ii},}\ in\ \emph{\bibinfo {booktitle} {Topics in Applied Physics}},\
  Vol.~\bibinfo {volume} {56},\ \bibinfo {editor} {edited by\ \bibinfo {editor}
  {\bibfnamefont{J.~D.}\ \bibnamefont{Joannopoulos}}\ and\ \bibinfo {editor}
  {\bibfnamefont{G.}~\bibnamefont{Lucovsky}}}\ (\bibinfo {publisher}
  {Springer},\ \bibinfo {address} {New York},\ \bibinfo {year} {1984})\
  p.~\bibinfo {pages} {61}%
  \bibAnnoteFile{NoStop}{Ley:1984}%
\bibitem{Lu:1996}%
  \BibitemOpen
  \bibfield{author}{%
  \bibinfo {author} {\bibfnamefont{Z.~H.}\ \bibnamefont{Lu}}, \bibinfo {author}
  {\bibfnamefont{D.~J.}\ \bibnamefont{Lockwood}},\ and\ \bibinfo {author}
  {\bibfnamefont{J.~M.}\ \bibnamefont{Baribeau}},\ }%
  \bibfield{journal}{%
  \bibinfo {journal} {Solid State Electronics}\ }%
  \textbf{\bibinfo {volume} {40}},\ \bibinfo {pages} {197} (\bibinfo {year}
  {1996})%
  \bibAnnoteFile{NoStop}{Lu:1996}%
\bibitem{Renmer:1973}%
  \BibitemOpen
  \bibfield{author}{%
  \bibinfo {author} {\bibfnamefont{O.}~\bibnamefont{Renmer}}\ and\ \bibinfo
  {author} {\bibfnamefont{J.}~\bibnamefont{Zemek}},\ }%
  \bibfield{journal}{%
  \bibinfo {journal} {Czech. J. Phys. B}\ }%
  \textbf{\bibinfo {volume} {23}},\ \bibinfo {pages} {1273} (\bibinfo {year}
  {1973})%
  \bibAnnoteFile{NoStop}{Renmer:1973}%
\bibitem{Kivelson:1979}%
  \BibitemOpen
  \bibfield{author}{%
  \bibinfo {author} {\bibfnamefont{S.}~\bibnamefont{Kivelson}}\ and\ \bibinfo
  {author} {\bibfnamefont{C.~D.}\ \bibnamefont{Gelatt}},\ }%
  \bibfield{journal}{%
  \bibinfo {journal} {Phys. Rev. B.}\ }%
  \textbf{\bibinfo {volume} {19}},\ \bibinfo {pages} {5160} (\bibinfo {year}
  {1979})%
  \bibAnnoteFile{NoStop}{Kivelson:1979}%
\bibitem{Singh:2002}%
  \BibitemOpen
  \bibfield{author}{%
  \bibinfo {author} {\bibfnamefont{J.}~\bibnamefont{Singh}},\ }%
  \bibfield{journal}{%
  \bibinfo {journal} {J. Non-Cryst. Solids}\ }%
  \textbf{\bibinfo {volume} {299}},\ \bibinfo {pages} {444} (\bibinfo {year}
  {2002})%
  \bibAnnoteFile{NoStop}{Singh:2002}%
\bibitem{Afanasev:2009}%
  \BibitemOpen
  \bibfield{author}{%
  \bibinfo {author} {\bibfnamefont{V.~V.}\ \bibnamefont{Afanas’ev}}, \bibinfo
  {author} {\bibfnamefont{M.}~\bibnamefont{Badylevich}}, \bibinfo {author}
  {\bibfnamefont{A.}~\bibnamefont{Stesmans}}, \bibinfo {author}
  {\bibfnamefont{A.}~\bibnamefont{Laha}}, \bibinfo {author}
  {\bibfnamefont{H.~J.}\ \bibnamefont{Osten}},\ and\ \bibinfo {author}
  {\bibfnamefont{A.}~\bibnamefont{Fissel}},\ }%
  \bibfield{journal}{%
  \bibinfo {journal} {Appl. Phys. Lett.}\ }%
  \textbf{\bibinfo {volume} {95}},\ \bibinfo {pages} {102107} (\bibinfo {year}
  {2009})%
  \bibAnnoteFile{NoStop}{Afanasev:2009}%
\bibitem{Seguini:2011}%
  \BibitemOpen
  \bibfield{author}{%
  \bibinfo {author} {\bibfnamefont{G.}~\bibnamefont{Seguini}}, \bibinfo
  {author} {\bibfnamefont{S.}~\bibnamefont{Schamm-Chardon}}, \bibinfo {author}
  {\bibfnamefont{P.}~\bibnamefont{Pellegrino}},\ and\ \bibinfo {author}
  {\bibfnamefont{M.}~\bibnamefont{Perego}},\ }%
  \bibfield{journal}{%
  \bibinfo {journal} {Appl. Phys. Lett.}\ }%
  \textbf{\bibinfo {volume} {99}},\ \bibinfo {pages} {082107} (\bibinfo {year}
  {2011})%
  \bibAnnoteFile{NoStop}{Seguini:2011}%
\end{thebibliography}%

\end{document}